\documentclass[a4paper]{jpconf}
\usepackage{graphicx}

\begin{document}

\title{High-Energy $\gamma$-ray Astronomy and String Theory}

\author{Nick E. Mavromatos}

\address{King's College London, Department of Physics, Strand, London WC2R 2LS, UK}

\ead{nikolaos.mavromatos@kcl.ac.uk}

\begin{abstract}

There have been observations, first from
the MAGIC Telescope (July 2005) and quite recently (September 2008) from the FERMI Satellite Telescope, on non-simultaneous arrival of
high-energy photons from distant celestial sources. In each case, the highest energy photons were delayed, as compared to their lower-energy counterparts, by clearly observable time intervals.
Although the astrophysics at the source of these energetic photons is still not understood, and such non simultaneous arrival might be due to non simultaneous emission as a result of conventional physics effects, nevertheless, rather surprisingly, the observed time delays can also fit excellently some
scenarios in quantum gravity, predicting Lorentz violating space-time ``foam'' backgrounds with a non-trivial subluminal vacuum refractive index suppressed linearly by a quantum gravity scale of order of the reduced Planck mass ($\sim 10^{18}$ GeV). In this talk, I discuss the MAGIC and FERMI findings in this context. First,
I review the high-energy astrophysics models for cosmic acceleration at celestial sources, stressing that currently there is no consensus as regards the observed delays. Then I derive estimates/bounds on the quantum gravity scale that reproduces the observed time delays on the assumption of a vacuum refractive index for photons with linear suppression, and argue on the consistency of such bounds with measurements from other Gamma Ray Telescopes, such as H.E.S.S. I then explain under which circumstances the MAGIC and FERMI findings could be accommodated in such models in agreement with all the other, currently available, astrophysics constraints of Lorentz Violation. The key features are: (i) transparency of the foam to electrons, (ii) absence of birefringence effects and (iii) a breakdown of the local effective lagrangian formalism. In contrast to other Quantum Gravity (field-theoretic) models with non-trivial optical properties available to date, a string model based on brane-worlds with the bulk space being punctured by space-time point-like D0-brane defects, that provide the seeds for Lorentz violating foamy structures, seems to respect all three requirements. The model provides an explanation for the observed photon time delays in a natural range of the string coupling and mass scale.
\end{abstract}

\section{Introduction and Summary: The MAGIC Observations and String Theory}

An alternative title for the talk could be \emph{MAGIC and String Theory}:
Usually the terminology \emph{M-theory}, with M standing for either \emph{Magical} or Marvelous or Mysterious, is attributed to the underlying (yet not completely understood) unifying theory of all known string theories~\cite{polch}, as a result of the many appealing duality and other symmetries it possesses, which result in the unification of the five known string theories, viewed as a low-energy limit of M-theory. This is a super-unification picture, which may prompt the way for a detailed understanding of the yet elusive theory of the quantum structure of space-time, otherwise termed as ``Quantum Gravity'' (QG). Our current knowledge/understanding of M-theory is limited. Nevertheless, for an analysis of some of the predictions of string theory that could have some relevance to observable low-energy physics this may not be an obstacle, as we shall attempt to discuss in this work.

\begin{figure}[ht]
\begin{center}
  \includegraphics[width=3cm, angle=0]{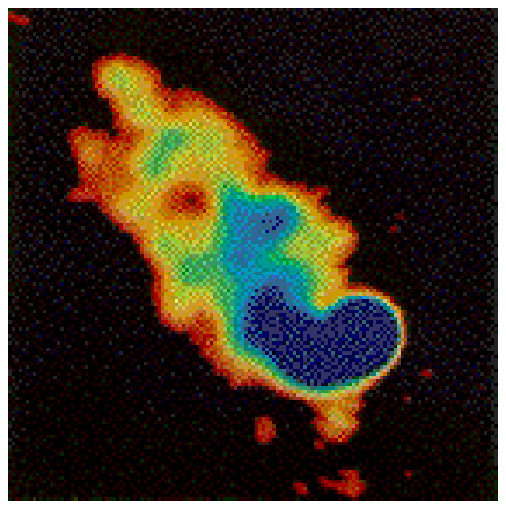} \hspace{2cm}
 \includegraphics[width=5cm,angle=0]{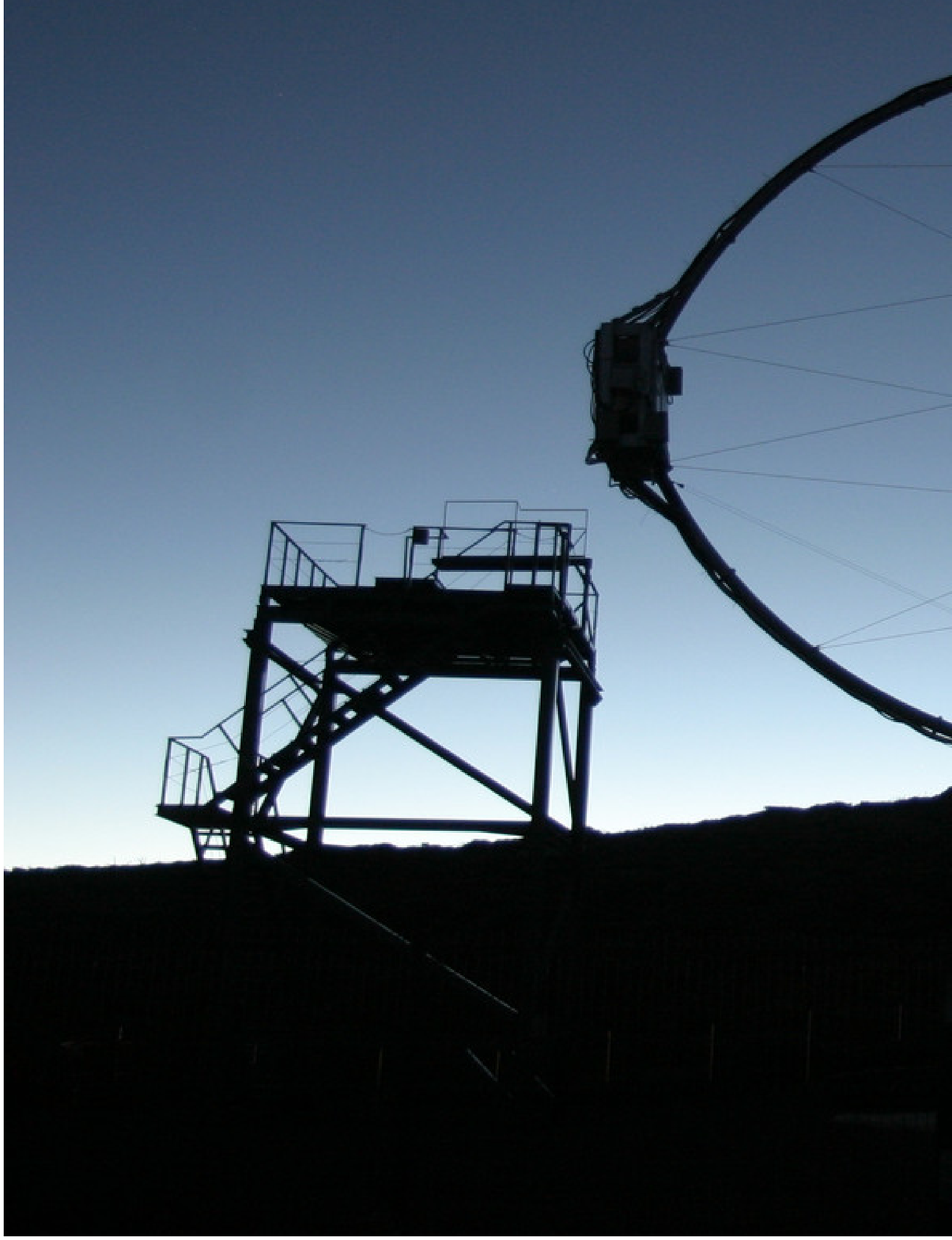} \vfill
\end{center}
\caption{Major Atmospheric Gamma-ray Imaging Cherenkov Telescope at the Canary Islands (Spain) Observatory (right panel). The telescope observed very high energy gamma rays from the active galactic nucleus Markarian 501 (radio image on left panel, by J.M. Wrobel and J.E. Konway,
picture taken from \texttt{http://www.vlba.nrao.edu/whatis/mark.html}) with energies up to the order of 10 TeV (i.e. $10^{12}$ eV).}
\label{fig:magic}
\end{figure}

In this review the word \emph{MAGIC}
is used for something completely different. It is
an acronym (\emph{M.A.G.I.C} = \emph{\textbf{M}}ajor \emph{\textbf{A}}tmospheric \emph{\textbf{G}}amma-ray \emph{\textbf{I}}maging \emph{\textbf{C}}herenkov telescope),
pertaining to the initials describing the full name of a Physics Instrument, specifically a Telescope based on the Canary Islands observatory (c.f. fig.~\ref{fig:magic}), dedicated to the study of
Cherenkov radiation emitted by highly energetic cosmic particles as they enter our atmosphere
From the study of the emitted Cherenkov radiation, when a highly energetic cosmic particle enters the Earth's atmosphere, which is characterised by a non-trivial refractive index, one can deduce several important conclusions on the nature of the initial particle and through this to try to understand the mechanisms of production of such energetic cosmic particles.

\begin{figure}[ht]
\begin{center}
\includegraphics[width=10cm]{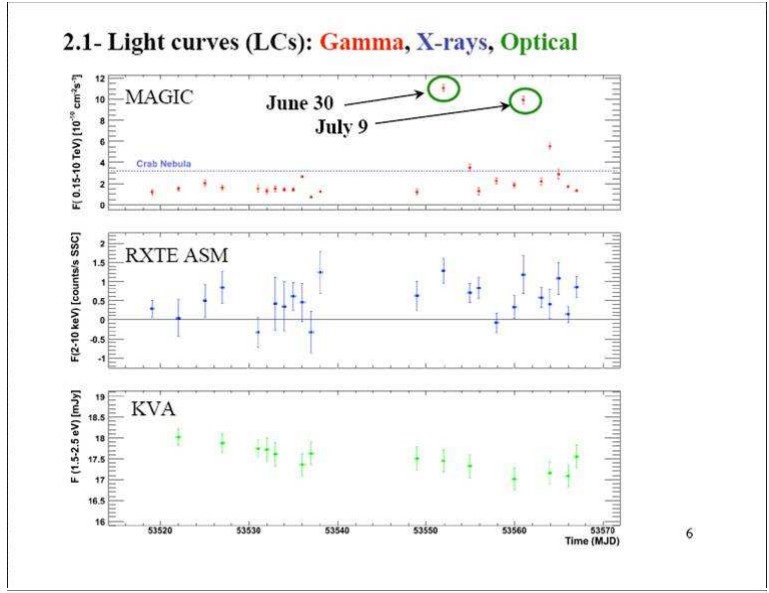}  \hfill \includegraphics[width=10cm]{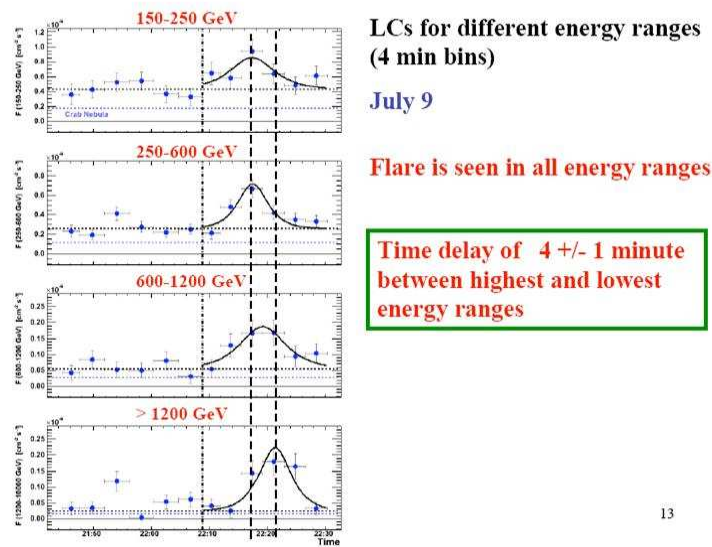}
\end{center}
\caption{ The observations of the MAGIC telescope~\cite{MAGIC} regarding very high energy Gamma Rays (with energies in the TeV range) showed that the most energetic photons
were delayed up to four minutes as compared with their lower-energy counterparts (in the 0.6 TeV or lower range). The figures show light curves (LC), i.e. the photon flux vs. time of arrival. Observations like this may be used to prompt new fundamental physics on the structure of space-time. The lower panel shows LC  at different energy ranges, demonstrating clearly the time delay (of order of 4 $\pm 1$ minutes) of the more energetic photons.}
\label{fig:magic2}
\end{figure}

On July 9th 2005, the telescope observed~\cite{MAGIC} (c.f. fig.~\ref{fig:magic}) very high energy gamma rays from the active galactic nucleus Markarian 501 (Mkn 501),
which lies at red-shift $z=0.034$  (i.e. about half a million light years away) from Earth,
with energies up to the order of 10 TeV  (1 TeV = $10^3$ GeV = $10^{12}$ eV), which were delayed up to four minutes as compared with their lower-energy counterparts (in the 0.6 TeV or lower range) (c.f. fig.~\ref{fig:magic2}). It was the first and currently the only observation
of such a distinct delay. The effect may be related to the astrophysics of the active galactic nucleus (\emph{source} effect), which, as we shall discuss below, is not well understood at present and hence there is no consensus among the astrophysicists on the appropriate mechanism for the production of such high-energy photons at the source.

These uncertainties prompted more ambitious, although admittedly looking far-fetched, explanations~\cite{MAGIC2},
pertaining to  new fundamental physics, affecting the photon \emph{propagation} due, for instance, to space-time \emph{foamy} vacuum structures that lead to modified dispersion relations for photons, that is
a departure from the Lorentz invariant energy (E)-momentum ($\vec p$) relations of Special Relativity, $E=|\vec p|c$.
Indeed, the reader should bear in mind that
at small length scales, of the order of the Planck length, $\ell_P = \sqrt{\frac{\hbar G_N}{c^3}} = \frac{\hbar}{M_P c} \sim 10^{-35}$~m, which is the characteristic scale that quantum gravity effects are expected to be dominant, the structure of space time may be quite different from what we perceive at our (low-energy) scales, and it might be even discrete and non-commutative, that is the space-time coordinates (as we perceive them at present) might be average values of non commutative quantum operators. Moreover, one may have highly curved non-trivial fluctuations of the space-time metric, giving space time a ``\emph{foamy}'' structure. In such complicated Quantum Gravity (QG) vacua, therefore, the very concept of Lorentz symmetry might break down at these short length scales, pointing towards the possibility of \emph{spontaneous Lorentz symmetry breaking} by the QG vacuum, since if Lorentz symmetry is intact,
it strictly prohibits such modifications in the dispersion relations. Such departures from the
 standard Special Relativity form of dispersion relations
correspond to the generation of a momentum dependent mass gap for the photon, and hence a non trivial
refractive index, as the photon propagates through the \emph{medium} of quantum gravity.
Here lies the origin of the \emph{spontaneous}  \emph{breaking} of the \emph{Lorentz symmetry}
by the ground state of such foamy quantum gravity models.

It is the point of this review to touch upon such issues, through the description of
of the most important physical consequences of such a breaking. As we shall see, surprisingly enough, many models of QG that entail such a breaking can already be falsified in current astrophysical experiments, which set very stringent bounds on Lorentz symmetry breaking. Some of the models we are going to discuss have experimental consequences that can be tested (or are already falsified(?)) in Nature, at least in principle. In this article we shall discuss in some detail only one class of theories of QG that can entail such a breaking, which is a subset of the modern version of string theory, including brane (domain-wall-like) defects in space time. Such defects will play the r\^ole of the non trivial space time structures that would be deemed responsible for the Lorentz symmetry spontaneous breaking by the ground state of these systems.

As we shall discuss here, there are very stringent conditions for such
exotic explanations of the MAGIC observations to come into play in agreement with
the plethora of many existing astrophysical tests of Lorentz symmetry.
The space-time foam must be transparent to electrically charged probes, such as electrons, while photons should exhibit non-trivial refractive indices in this theory, but with no birefringence effects.
Moreover, the local effective lagrangian description of the effects, that is a low-energy representation
of the QG medium effects in terms of local higher derivative operators in flat space-time backgrounds,
should break down.

Thus, although one cannot exclude the possibility that
\emph{both} effects, source and propagation due to quantum gravity, may be simultaneously responsible for the observed photon delays in the MAGIC experiment, nevertheless the available theoretical models that do the job are very limited.
The thesis of this article will be that only certain models~\cite{emnnewuncert} of (the modern version of) string theory, including space-time defects, whose dynamics breaks Lorentz symmetry and provides a ``foamy'' structure of space-time, can offer an explanation for the MAGIC photon-arrival-times anomaly, which actually might be unique in interpreting these observations as a result of a stringy space-time foam situation, in a way consistent with all the current astrophysical constraints on Lorentz symmetry.

The key features of such stringy models are: (i) transparency of the foam to electrons, (ii) absence of birefringence effects and (iii) a breakdown of the local effective lagrangian formalism. In contrast to other Quantum Gravity (field-theoretic) models with non-trivial optical properties available to date, a string model based on brane-worlds with the bulk space being punctured by space-time point-like D0-brane defects, that provide the seeds for Lorentz violating foamy structures, seems to respect all three requirements. Due to electric charge conservation, only electrical neutral excitations (such as photons) interact with the D-particle defects, thereby leading to time delays proportional to the photon energies, but with no birefringence effects, in the sense that the delays are independent of the photon polarization.
Moreover, the recoil and the quantum fluctuations of the D-particles, during their scattering with matter excitations,
imply a breakdown of the local effective lagrangian formalism in the sense that the phenomena cannot be wholly represented
by the addition of local higher-derivative operators in a flat-space time effective lagrangian.
The string model thus provides an explanation for the observed photon time delays, in a natural range of the string coupling and mass scale, and avoids all the other stringent constraints of Lorentz Invariance Violation coming from non observation of birefringence effects or of very high energy photons, that would be the case if local effective lagrangian formalism were in place, as a result of the modification of the energy thresholds for pair production  in the scattering of ultra-high-energy photons ($\sim 10^{19}$ eV) with infrared ones in the Universe~\cite{sigl}.

A word of caution to the reader is due at this point. Our point is not to advocate string theory as a superior candidate to other quantum gravity models available to date, but rather to discuss situations in which string theory predictions can be falsified by experiment. And high energy photon astrophysics may be a useful arena for this purpose!
Of course, it goes without saying that, at present, we are very far from reaching any conclusions on such matters. Definite falsification of these stringy quantum gravity scenarios, if at all possible,  would require a plethora of further studies by means of other high-energy astrophysics processes and observations.

To understand in detail the possible explanations of the MAGIC ``photon anomaly'', and why string theory comes into play as a potential explanation provider, I will  first recapitulate the up to date
knowledge on theoretical models of QG, followed by a brief review of the scenarios for the production of very high energy gamma rays in the Universe, known to date. In this way I hope I will be able to
convince the reader that the theoretical uncertainties in both fields, astrophysical production of high energy photons and possible propagation effects due to quantum gravity media, are more or less on equal footing, and presently there is no consensus in both communities. This leaves the field open to speculations and it
is in this sense that the exotic explanation of the MAGIC effect we shall provide here, based on string theory, may not sound so far fetched after all! Hopefully, at the end of the day, the reader will be able to make his/her own judgement on whether such a theory is really required to explain the facts.

And such facts, at present, seem to be increasing in numbers.
Indeed, three years after the MAGIC observations, in September 2008, the FERMI (formerly known as GLAST) Satellite Telescope~\cite{grbglast}, also observed time delays of the higher-energy photons, from the distant Gamma Ray Burst (GRB) 080916c. As we shall discuss in this talk, the pattern of the delay fits~\cite{emnnewuncert} the above-mentioned string model of quantum-gravity-induced refractive index, with the pertinent quantum gravity scale being essentially the same as that inferred from the MAGIC observations. Viewed as a lower bound, this scale is also compatible with that obtained from other Gamma-Ray data of the H.E.S.S. Collaboration~\cite{hessnew}.

The structure of the article is as follows: in the next section, \ref{sec:LV}, I discuss issues pertaining to the violation of Lorentz invariance in media or
quantum electrodynamics vacua with non trivial structure that break Lorentz invariance, in the sense of inducing a non-trivial vacuum refractive index, \emph{e.g}. the case of thermal plasma. Such cases may be thought of as analogues of the Quantum Gravity (QG) space-time foam vacuum that may also not respect Lorentz symmetry but can also characterise the source regions of the cosmic high energy Gamma Rays. In this latter sense, it is important to understand photon propagation in those cases first, so as to disentangle possible source effects from the Quantum-Gravity-induced ones. In section \ref{sec:3}, we discuss possible interpretations of the MAGIC (and FERMI) observations, including exotic ones involving QG dispersive media. We discuss bounds and sensitivities of various astrophysical experiments, and state carefully the stringent requirements that must be met by a theoretical model of QG, in order for the observed delays in the MAGIC and FERMI Telescopes to be attributed to effects due to a space-time-foam medium, in agreement with all other current tests. All these requirements are surprisingly respected by a model in the modern version of string theory, which is discussed in section \ref{sec:string}. The model involves membrane-like defects in a higher-than-four dimensional space time,
with our world being viewed as a hyper-membrane (D(irichlet)-brane) embedded in this space time.
The model is of the kind of large-extra-dimension models to be tested at LHC and future colliders, with the important ingredient of having point-like space-time (D-particle) defects, responsible for the ``foamy'' structure of space time. Finally, section \ref{sec:5} contains our concluding remarks.

A note is in order at this point concerning the \emph{units} used in this work: throughout the article, unless otherwise stated, we shall work in \emph{natural Planck units}, in which $\hbar = c = 1$. In these units, length and time are identified, and they are inversely proportional to mass or energy. The
 latter are also identified and expressed in units of multiples of eV (=$1.6 \times 10^{-19}$ Joules), in particular GeV (=$10^9$ eV) and TeV (=$10^{12}$ eV) in this work. From time to time, for concreteness, the speed of light \emph{in vacuo} $c$ and Planck's constant $h$ or $\hbar =h/2\pi$ may appear explicitly in some formulae.

\section{Lorentz Invariance and the Vacuum Structure of Quantum Fields \label{sec:LV}}

One of the cornerstones of Modern Physics is Einstein's theory of Special Relativity (SR), which is based on the assumption that the speed of light in vacuo $c$ is an invariant under all observers, and in fact this implies the Lorentz transformation in flat space times, and the nature of $c$ as a limiting velocity for \emph{all} particle species in SR.

The generalization (by Einstein) of SR to include curved space times, that is the theory of General Relativity (GR), encompasses SR locally in the sense of the \emph{strong form of the equivalence principle}. According to it,  \emph{at every space-time point, in an arbitrary gravitational
field, it is possible to choose a locally
inertial (`free-float')  coordinate frame, such that within a sufficiently
small region of space and time around the point in question,
the laws of Nature are described by special relativity, i.e. are of the
same form as in unaccelerated Cartesian coordinate frames in the absence
of Gravitation.} In other words, locally one can always make a coordinate transformation
such that the space time looks {\it flat}. This is not true globally, of course, and this is why GR is a more general theory to describe gravitation. The equivalence principle relies on another fundamental invariance of GR, that of general coordinate, that is the invariance of the gravitational action under arbitrary changes of coordinates. This allows GR to be expressed in a generally covariant form.

In such a locally \emph{Lorentz-invariant} vacuum,
the photon dispersion relation, that is a local in space-time relation between the photon's four-wavevector components $k^\mu =(\omega, \vec k)$ (where $\omega$ denotes the frequency, and $\vec k$ the momentum) reads in a covariant notation:
\begin{equation}
  k^\mu k^\nu \eta_{\mu\nu} = 0
\label{photonflat}
\end{equation}
where repeated indices $\mu, \nu = 0, 1, \dots 3$, with $0$ referring to temporal components, denote summation and $\eta_{\mu\nu}$ denotes the Minkowski space-time metric, with components $\eta_{00} = -1, ~\eta_{0i}=\eta_{i0}=0, ~ \eta_{ij}=\eta_{ji}=\delta_{ij}~, i = 1,2,3$ with $\delta_{ij}$ the Kronecker delta symbol.

The above relation (\ref{photonflat}) implies the equality of all three kinds of photon velocities \emph{in vacuo} that stem from its wave nature (due to the particle-wave duality relation):
\begin{eqnarray}
&& {\rm phase}: \qquad v_{\rm ph} = \frac{\omega}{|\vec k|} \equiv \frac{c}{n(\omega)} = c \nonumber \\
&& {\rm group}: \qquad v_{\rm gr} = \frac{\partial \omega }{\partial |\vec k|} \equiv \frac{c}{n_{\rm gr}(\omega)} = c~,~ \quad n_{\rm gr}(\omega) = n(\omega) + \omega \frac{\partial n(\omega)}{\partial \omega}  \nonumber \\
&& {\rm front}: \qquad v_{\rm front} = c/n(\infty) = c
\label{velocities}
\end{eqnarray}
since the phase and group \emph{refractive indices} of the trivial \emph{vacuum } equal unity $n(\omega) = n_{\rm gr}(\omega) = 1$. For brevity we shall work from now on in units where $c=1$.

\subsection{Photon Propagation in Conventional Media  and in Non-trivial field-theory vacua with a refractive index \label{sec:ntv}}

The above results (\ref{velocities}) changes significantly when light propagates in a material \emph{medium}, in which its speed is different from $c$ in vacuo. This is due to the non trivial refractive index the material has, as a result of the electromagnetic interactions of the photon with the electrons in the medium. As we shall discuss later on, this case seems to bear some quite instructive analogies with our string model of space time foam, which lies at the focus of our attention here.

The simple model of quantum oscillators has been adopted by Feynman~\cite{feynman} as a
simplified but well motivated analogue for describing the situation in  the case of
photons in ordinary media.
In that case, the electrons of the medium, of mass $m$, are represented by simple harmonic oscillators, with frequency $\omega_0$,
which provides the necessary \emph{restoring force} during the scattering of light off the electrons in order to keep the latter oscillating around their initial position.
In that problem the induced refractive index is obtained by the reduction of the phase velocity of the photon wave.

It can then be shown by means of an elementary analysis, based on quantum mechanics~\cite{feynman},
that light propagates through the medium with a
speed $c/n$, where $n$ is the refractive index of the medium,
which is given by the following fomrmua|:
\begin{equation}
   n = 1 + \frac{{\rho_e}_e e^2 }{2\epsilon_0 m (\omega_0^2 - \omega^2)}.
   \label{refrordinary}
   \end{equation}
In this relationship, $\epsilon_0$ is the dielectric constant of the vacuum and $n_e$ is the area density of
electrons in the medium (plate in the example of \cite{feynman} adopted here for concreteness), which is given by $n_e = \rho_e \Delta z$,
where $\rho_e$ is the volume density of electrons.

Thus causes a delay $\Delta t$ in traversing the distance $\Delta z$, given by:
$\Delta t = (n - 1)\Delta z/c$.
We see in (\ref{refrordinary}) that the refractive index in an ordinary medium is inversely proportional to (the square of) the frequency $\omega$ of light, as long as it smaller than the oscillator
frequency, where the refractive index diverges.

If the couplings of the two polarizations of the photon to the electrons in the medium are different,
the phenomenon of birefringence emerges, namely different refractive indices for the two
polarizations. Moreover, we see from (\ref{refrordinary}) that the propagation of light is
subluminal if the frequency (energy) of the photon $\omega < \omega_0$, whereas it is
superluminal for higher frequencies (energies)~\cite{feynman}.
This reflects the fact that the phase shift induced for the scattered light can be either positive or negative, but there such a superluminal refractive index causes no issue with causality,
since the speed at which information may be sent is still subluminal.

As we shall see later on, this conventional model will provide us with a rather good analogue of what happens in some string models of quantum space time foam, also characterized by non-trivial refractive indices, which we shall analyse in section \ref{sec:string}.
However, as we shall see there, contrary to the conventional situation discussed in this section, in the string case the refractive index is found proportional to the photon frequency, while the effective mass scale that suppresses the effect is the quantum gravity (string) scale and not the electron mass as in (\ref{refrordinary}).

\begin{figure}[ht]
 \begin{center} \includegraphics[width=9cm]{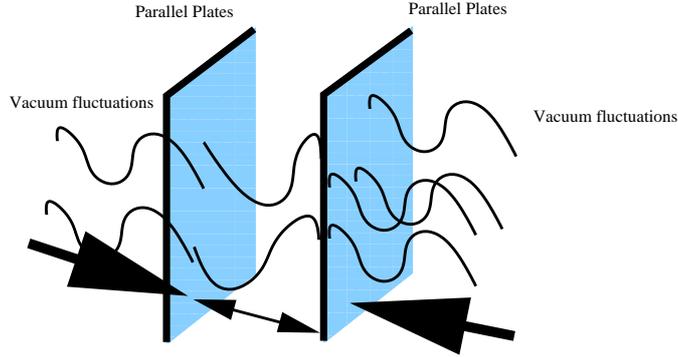} \end{center}
\caption{The Casimir Vacuum of Quantum Electrodynamics: in the compact space between the parallel plates (or more generally in other compact geometries), the quantum fluctuations of the electromagnetic field are responsible for the macroscopically measured
Casimir force between the plates, scaling with the size $L$ of the compact space as $L^{-4}$ in four space-time dimensions. The force can be either attractive or repulsive, depending
on the geometric set up. However, besides this force from ``nothing'', the Casimir Vacuum provides another interesting phenomenon in quantum physics. According to calculations by Scharnhorst and collaborators~\cite{scharn}, the presence of the boundaries
breaks manifestly Lorentz invariance, and results in
modified dispersion relation, and hence a non-trivial refractive index, for the virtual photons
of the \emph{non-trivial Quantum Electrodynamics vacuum} in the compact region. From this latter perspective, the situation is entirely analogous to the what happens in thermal plasmas, such as the ones occurring in the interior of stars or source regions of active galactic nuclei, which we are interested in in this work. There, the periodic boundary conditions of the Casimir case are replaced by similar ones due to the finite temperature, which also break Lorentz symmetry of the ground state.}
\label{fig:casimir}
\end{figure}

The r\^ole of a non-trivial refractive index material can be played under certain circumstances by
a \emph{non-trivial vacuum} in which photons propagate, such as quantum electrodynamics at finite temperature plasmas~\cite{latorre} or the Casimir vacuum between parallel capacitor plates~\cite{scharn} (or other geometries, as long as the space is bounded appropriately)
(c.f. fig.~\ref{fig:casimir}). In such cases, the loop corrections due to \emph{vacuum polarization}, \emph{i.e.} creation and annihilation of virtual electron-positron pairs, in quantum electrodynamics (QED) result in a modified photon propagator, and a non-trivial group velocity and refractive index $n(\omega) \ne 1$. The reason for this is the breaking of Lorentz invariance, due to either the existence of spatial boundaries (Casimir vacuum) or finite temperature (thermal vacuum in case of plasmas). Moreover one may consider quantum electrodynamics in a homogeneous and isotropic Friedman-Robertson-Walker (FRW) expanding-Universe background~\cite{hath} and examine the non-trivial effects of vacuum polarization on photon propagation there. This was in fact historically the first instance where the effects of curvature induced \emph{superluminal} propagation for low-frequency photon modes.

The situation concerning all the above cases can in fact be represented in a rather unified way, covering all cases, by the following formula giving the group velocity of (low-energy) photons in such non-trivial vacua due to vacuum polarization in four-dimensional QED~\cite{latorre}:
\begin{equation}
v_{\rm gr} = 1 - \frac{44}{135}\alpha^2 \frac{\rho}{m_{\rm e}^4}
\label{unified}
\end{equation}
where  $\rho$ is the energy density \emph{relative} to the standard vacuum. This formula is valid in all cases except the Gravitational Background case of \cite{hath}, where $\alpha^2$ should be replaced by $\alpha m_{\rm e}^2 G_{\rm N}$, with $G_N$ the Newton gravitational constant.
In the (four space-time dimensional) Casimir vacuum, for instance, the presence of boundaries in space (e.g. capacitor parallel plates at a distance $L$ in the simplest geometry), imply loss of low-energy photon modes, and as such the energy density of the vacuum is lowered relative to the standard one (without the boundaries). In this case $\rho = - \frac{\pi^2}{720 L^4} < 0$ in (\ref{unified}), and one has the phenomenon that the effective low-energy photon modes appear to propagate in a \emph{superluminal way}, $v{\rm gr} > 1$ (we give here the formula for propagation \emph{perpendicular} to the plates for concreteness and ease of comparison with the plasma case later on):
\begin{equation}
v_{\rm gr}^{\rm Casimir} = 1 + \frac{11 \pi^2}{8100}\alpha^2 \frac{1}{L^4 m_{\rm e}^4} > 1
 \label{casimir}
 \end{equation}
 But as explained in \cite{latorre}, and mentioned above, there is no contradiction with relativity here, as this result applies only to low energy photons, with energies much lower than $m_{\rm e}$, and indicates the effective loss of degrees of freedom due to the spatial boundaries.

In the plasma case at finite temperature $T$, which can characterise the interior regions of stars or
active galactic nuclei, of interest to us in this work,
there is a formal analogy~\cite{latorre} between the r\^oles played by the temperature $T$ and the plate separation $L$. They both break Lorentz invariance in a rather similar way. In fact, there is a correspondence between the Casimir and thermal-plasma vacuum formulae by replacing $2T $ by $L^{-1}$. In the plasma case, the low energy photon modes with momentum $k^2 \ll m_{\rm e}^2$ ($k \equiv |\vec k|$) have a group velocity:
\begin{equation}
v_{\rm gr}(kT \ll m_{\rm e}^2 ) = 1 + \frac{11 \pi^2}{8100}\alpha^2 \left(\frac{2T}{m_{\rm e}} \right)^4 ~> ~1
\end{equation}
which can be directly compared with the Casimir one (\ref{casimir}) upon the substitution $2T \to L^{-1}$.

For future reference we give the group velocity for the high-energy (as compared to the effective QED scale $m_{\rm e}$) photon modes~\cite{latorre} in the finite $T$ plasma case:
\begin{equation}
v_{\rm gr}(kT \gg m_{\rm e}^2 ) = 1 - \frac{\alpha^2 }{6}\left(\frac{T}{k}\right)^2 {\rm ln}^2\left(\frac{kT}{m_{\rm e}^2}\right) ~< ~1
\label{highenergy}
\end{equation}
which again can be compared with the corresponding Casimir vacuum case (\ref{casimir}), for photon modes \emph{perpendicular} to the plates~\cite{scharn}, upon the replacement $2t \to L^{-1}$.
Similar effects also characterise the curved background case of \cite{hath}, where again the high-momentum photon modes are subluminal.
We remark at this point that the relation (\ref{highenergy}) can characterise the source regions of active galactic nuclei, and thus
such effects can be responsible for inducing time delays of photons from these regions.
We shall come back to it, when discussing the MAGIC delays in section \ref{sec:3magicqg} below.

We notice here the momentum $(k)$ dependence of the subluminal velocity for the high-energy photon modes.
In such non-trivial vacua, therefore, high energy photon modes will exhibit a momentum dependent subluminal refractive index, which will affect their arrival times at the observation point, if one considers simultaneous emission of modes within a certain energy range. From (\ref{highenergy}) we observe that the higher the momentum $k$ of the high-energy photon mode the higher the group velocity, since $\frac{\partial v_{\rm gr}(kT \gg m_{\rm e}^2)}{\partial k} = \frac{\alpha^2 T^2}{3k^3}\left({\rm ln}(\frac{kT}{m_{\rm e}^2})\right) > 0$, and thus fast modes will arrive first if emitted simultaneously in this vacuum.

At this point it is instructive to make some clarifications regarding the speed of light and causality, that is the fact that signals do not arrive before they occur.
The phase, front and group velocities
have all been found to exceed the value of the speed of light in vacuo. However, this is not in conflict with
causality, that is the fact that signals never arrive before they occur.

Indeed, the various light velocities do not have to be subluminal, as they carry no information.  Information can be transmitted by (or not) sending pulses. The information then \emph{appears} at first sight to propagate with the group velocity, \emph{i.e.} the velocity of the peaks of the pulses. However,
as experimentally demonstrated in several instances since the 1980's, the group velocity of the photons can also be super-luminal. This is because the group velocity can be disentangled from the \emph{signal velocity}, that is the speed by which information can be transferred, and thus it can exceed the speed of light \emph{in vacuo}, $c$, without contradicting the \emph{causality} requirements of Special Relativity. Group velocities larger than $c$ can occur \emph{e.g}. in tunneling experiments and appear to lead to superluminal transmission. However there is no
contradiction with causality or Special Relativity in such cases. The error lies in identifying the peak of the pulse with the temporal position of the carried information. For example, a Gaussian-shaped pulse can be detected long before its peak due to the rise of intensity at earlier times. Therefore, a different kind of signal must be considered, where no information at all is sent out before a certain moment of time. For such signals, it can be proven that the earliest time at which that switching event can be observed is limited exactly by propagation with the vacuum velocity $c$. A so-called \emph{precursor} is traveling with that speed, but is normally too weak to be detected, except in certain circumstances.
For direct measurements of optical precursors in regions with anomalous dispersion the reader is referred, for instance, to ref.~\cite{precur}.

Super-luminal group photon velocities have been measured in laser pulses passing through specially prepared materials in ref. \cite{brunner}. Using special set up involving optical fibers, a super-luminal  group velocity of photons (in the fiber) has been measured and found different from the \emph{signal
velocity}, which is defined as the speed of the front of a square pulse. It is the signal speed that
cannot exceed $c$, due to \emph{causality}, as mentioned above. The precursors (or ``forerunners'') of the signal, mentioned above, which travel with sub-luminal speed, arrive first before the main front. This signal velocity
was measured for the first time directly in an experiment in \cite{brunner}, and indeed was found to be less than $c$.

This is an important issue that the reader should have in mind, especially when we discuss non-trivial vacuum refractive indices in several
non trivial ground states of quantum systems, including gravity.

In fact, as argued in ref.~\cite{libe}, where we refer the interested
reader for details, specifically for the case of Casimir vacuum~\cite{scharn} but the discussion can be generalised, super-luminal group velocities are ``benign'' as far as causality and  compatibility with the kinematics of Special Relativity are concerned. In particular, as stressed in \cite{libe}, kinematics of Special Relativity requires \emph{only} an invariant speed \emph{not} actually a maximum one. Moreover, causality can be guaranteed in such super-luminal cases because the pertinent kinematics is equivalent to an ``effective'' metric in space-time, $g_{\rm eff}^{\mu\nu}$, $\mu,\nu =0,\dots 3$, different from the Minkowski one, that describes the kinematics in the non-trivial (Casimir) vacuum. The modified dispersion relations leading to super-luminal group velocities (\ref{casimir}) acquire the form
 \begin{equation}
  k_\mu k_\nu g_{\rm eff}^{\mu\nu} = 0~,
  \label{effmetr}
\end{equation}
where as usual repeated indices denote summation, $k_0 = -\omega$, $k_i = (\vec k)_i, i=1,2,3$,
$g_{\rm eff}^{\mu\nu} = \eta_{\mu\nu} + \xi n^\mu n^\nu $, with $\eta^{\mu\nu}$
the (inverse) Minkowski metric, $n^\mu$ a unit (space-like) vector
orthogonal to the plates of fig.~\ref{fig:casimir}, and $\xi = \frac{11 \pi^2 \alpha ^2}{4050 L^4 m_{\rm e}^4}$, with $L$ the distance between the Casimir plates.

The basic point of the discussion in \cite{libe}, and how causality is maintained, is that
the presence of the effective metric in (\ref{effmetr}) widens slightly (due to the deviations of order $\xi \ll 1$ of the effective metric from the Minkowski one) the light cone in the direction orthogonal to the plates,
and hence light in that direction travels at a speed $c_{\rm light} > c$, while light in the direction parallel to the plates still travels with speed $c$. In a given inertial reference frame, moving with  four-velocity $u^\mu$ with respect to the rest frame of the apparatus of fig.~\ref{fig:casimir},
the photons inside the Casimir cavity travel, at a direction perpendicular to the plates, with a definite speed $c_{\rm light}^{(u)} > c$, which has \emph{only one value} for each observer and is not universal among observers.
In this way  super-luminal
group velocities (\ref{casimir}) are compatible with causality, since violation of the latter occurs \emph{only } if signals travel with the \emph{same} speed greater than $c$ in two different frames.
The reader should have these features in mind when studying space-time foam theories of quantum gravity.

\subsection{Non-trivial Optical properties of the Quantum-Gravity Vacuum ? }

A truly unspeakable feature on the speed of light may appear when considering the ground state of \emph{Quantum Gravity} \emph{per se } as a non-trivial vacuum with non-standard optical properties, leading to a \emph{non-trivial} refractive index. This may characterise, for instance, certain approaches in which path-integration over microscopic singular fluctuations of the metric field, such as Planck size black holes and other topologically non-trivial configurations, implies a sort of ``\emph{foamy}'' structure of the space time at small (Planck-size) length scales, over which photons can propagate,
a proposal made initially by J.A. Wheeler~\cite{wheeler}.
In such situations, the concept of a \emph{local Effective Quantum-Field Theory} Lagrangian may \emph{break down}, and the situation resembles that of a quantum-decoherent motion of matter in open quantum mechanical systems interacting with an environment~\cite{hawking,ehns,banks}.

 The ground state of such QG foam situations
may behave as a (subluminal) \emph{refractive medium}, as suggested originally in \cite{aemn},
which to our knowledge, constitutes the first concrete attempt to consider the non-trivial optics effects effect of such vacua on massless particle (photon) propagation,
based on earlier works by the current authors on non-critical strings in black hole backgrounds~\cite{emn}.
We note at this stage that similar ideas on modified dispersion relations for particles in a quantum gravity medium, but on a purely phenomenological basis, without any attempt to present concrete models, have also been advocated in ref.~\cite{mestres}. In the quantum gravity case, the effective modified dispersion relations of the low-energy matter theory, assumes the generic form:
\begin{equation}
     E^2 = |\vec p|^2 + m^2 + \sum_{n=1}^{\infty} c_n |\vec p|^2\left(\frac{|\vec p|}{M_{\rm QG}}\right)^n
\label{mdr}
\end{equation}
where $m$ is the rest mass of the probe, and $c_n$ are constant coefficients, with signature and values that depends crucially on the type of theory considered. It is not clear whether the series converges, or is resummable, as this information depends crucially on the details of the underlying microscopic theory.
The important point to notice in (\ref{mdr}) is that the natural suppression scale of the (Lorentz-symmetry-violating) correction terms of the standard Special Relativity dispersion relation is that of Quantum Gravity itself, $M_{\rm QG}$. According to what was mentioned earlier, there seems to be no fundamental issue with violation of (micro-)causality~\cite{libe} or contradiction with the principles of Special Relativity in cases where super-luminal group velocities arise from the QG anomalous dispersion relations (\ref{mdr}). The important point is to observe that effective metrics, as in the Casimir
vacuum case (\ref{effmetr}), can indeed be found in order to describe the QG anomalous dispersion
(\ref{mdr}) at least in some cases, that we shall be interested in here. Thus causality can be saved for those QG cases by applying
the same logic as for the Casimir vacuum, discussed previously.
Nevertheless, as we shall discuss later on in subsection \ref{sec:biref}, in cases where super-luminal modes exist, one has QG-induced birefringence phenomena, which are severely constrained by the current phenomenology (\emph{i.e} by the absence of the relevant signals).

Before closing this subsection we also mention that there is another approach towards modified dispersion relations of particles, of the type (\ref{mdr}), based on the so-called doubly special or deformed special relativity  (DSR) theories~\cite{dsr,smolin}. In such approaches, which are formulated on flat space times, the modification to the dispersion relation arises by the postulate that the local symmetry group of space time is no longer the Lorentz group but a different one. In their original version~\cite{dsr}, DSR modified dispersion relations were obtained by the requirement that the length scale of quantum gravity (``Planck'') remain invariant under transformations, which thus leads to deviations from the Lorentz group. There is no unique prescription to achieve this, however, and in this way one may even arrive at models where there is no upper limit in momentum. In their subsequent version~\cite{smolin}, DSR models postulated the existence of upper limits in velocities of species, and in this way the modified local group was determined by appropriate combinations of dilatations and translations.

It is unclear to us whether such theories are fundamental or effective, and most likely, as fundamental there will be not consistent in their quantization. For instance, at present a known problem is the behaviour of multi-particle states in such DSR fundamental theories. We shall not discuss these theories further here, apart from mentioning that, as in the stringy model we shall consider in section \ref{sec:string}~\cite{aemn,horizons,recoil}, they are not characterised by birefringence effects, leading only to subluminal propagation (defined appropriately~\cite{dsr,smolin}). There is, however, an important difference from the stringy model: the action of gravity is universal among particle species in DSR theories, and therefore the resulting modification in the dispersion relation pertain to all species. In contrast, for specifically stringy reasons to be outlined in section \ref{sec:string}, in the string model only photons and at most electrically neutral particles~\cite{ems} are allowed to interact non-trivially with the stringy defects, and are thus subjected to non-trivial modification of their dispersion relations.

\section{Possible Interpretations of the MAGIC Effect \label{sec:3}}

To understand in detail the possible explanations of the MAGIC observations, let us first recapitulate the up to date knowledge on the production of very high energy gamma rays in the Universe~\cite{deangelis} and ways of cosmic acceleration. I must stress that there is still no consensus among the astrophysical community as regards the various ways of particle acceleration at various regions in the Universe, and in fact it is most likely that there are several mechanisms taking place, depending on the source. The MAGIC observations added to this puzzle, in particular why the more energetic photons from the Mk501 Galaxy
arrived later than the lower energy ones. This is the only case up to now where such delays have been observed. The situation i n our understanding of the source mechanism for the production of high energy particles from cosmic sources will improve only by making more and more precision measurements from a variety of celestial sources, which is currently under way, through several terrestrial and extra-terrestrial facilities on high-energy astrophysics, which are either under construction or just have been put in operation.

It is these uncertainties in the conventional astrophysics of the sources that allow for speculations that
fundamental physics, such as photon propagation in a quantum gravity ``medium'', might play a significant r\^ole on the MAGIC effect. However, if the latter has a chance of being true it must respect all the other stringent astrophysical constraints on Lorentz invariance that are currently available. As we shall discuss in section \ref{sec:string}, it seems that, at present, only a specific string theory model of quantum foam, in which only photons are not transparent to the foam effects,  could stand up to this chance.

\subsection{Conventional Astrophysics mechanisms for cosmic Very-High-Energy (VHE) Gamma-Ray production}

Gamma rays constitute the most interesting part of the spectrum of active galactic nuclei (AGN).
An AGNs is a compact region at the centre of a galaxy, with much higher than normal luminosity over some or all of the electromagnetic spectrum (in the radio, infrared, optical, ultra-violet, X-ray and/or gamma ray wavebands). The radiation from AGN is believed to be a result of accretion on to the supermassive black hole at the centre of the host galaxy. AGN are the most luminous persistent sources of electromagnetic radiation in the universe, and as such can be used as a means of discovering distant objects; their evolution as a function of cosmic time also provides constraints on cosmological models.

Gamma Rays with energies higher than 20 MeV and up to TeV have been observed today from such AGNs.
It is customary (although somewhat arbitrary) to classify these Gamma Rays according to their energies as follows:
\begin{itemize}

\item{(i)} \emph{High-Energy} Gamma rays: with energies from 20 MeV - 100 GeV~,
\item{(ii)} \emph{Very High-Energy} Gamma rays: with energies from 100 GeV - 30 TeV~,
\item{(iii)} \emph{Ultra High-Energy} Gamma rays: with energies from 30 GeV - 30 PeV~,
\item{(iv)} \emph{Extremely High-Energy} Gamma rays: with energies from 30 PeV - ?.
\end{itemize}
Theoretically, the last category incorporate energies up to the ultraviolet cutoff energy scale
(Planck-scale energies $10^{19}$ GeV) that defines the structure of low-energy field theories as we know them.

The production of very high energy gamma rays is still not understood well, and constitutes a forefront of research on galactic and/or extragalactic astrophysics.

\begin{figure}[ht]
\begin{center}
  \includegraphics[width=5cm]{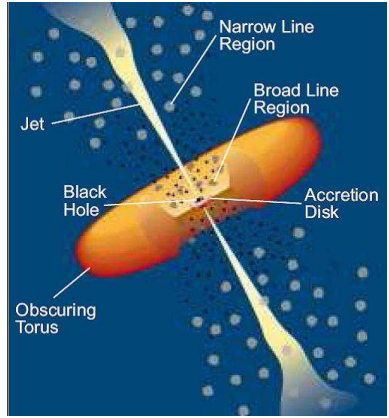}
\end{center}
\caption{The basics of a cosmic accelerator model (picture taken from ref.~\cite{agngravcoll}): Very high energy Photon production from Gravitational Energy conversion (Penrose process) in AGNs, believed to take place in AGN Mkn501.
Relativistic matter, such as electrons are beam ejected as a result of the enormous gravitational energy available during the collapse process forming a black hole at the center of the galaxy.
Then such electrons undergo synchrotron radiation due to their interaction with the magnetic fields existing in the galactic region, and eventually inverse Compton scattering (IC), i.e. interactions of these very high energy electrons with low-energy photons (say of e energies), to produce
TeV photons from Mkn501 observed by MAGIC (see fig.~\ref{fig:ssc}). This combined process is called Synchrotron-self-Compton process.}
\label{fig:crproduction}
\end{figure}

\begin{figure}[ht]
\begin{center}
  \includegraphics[width=7cm]{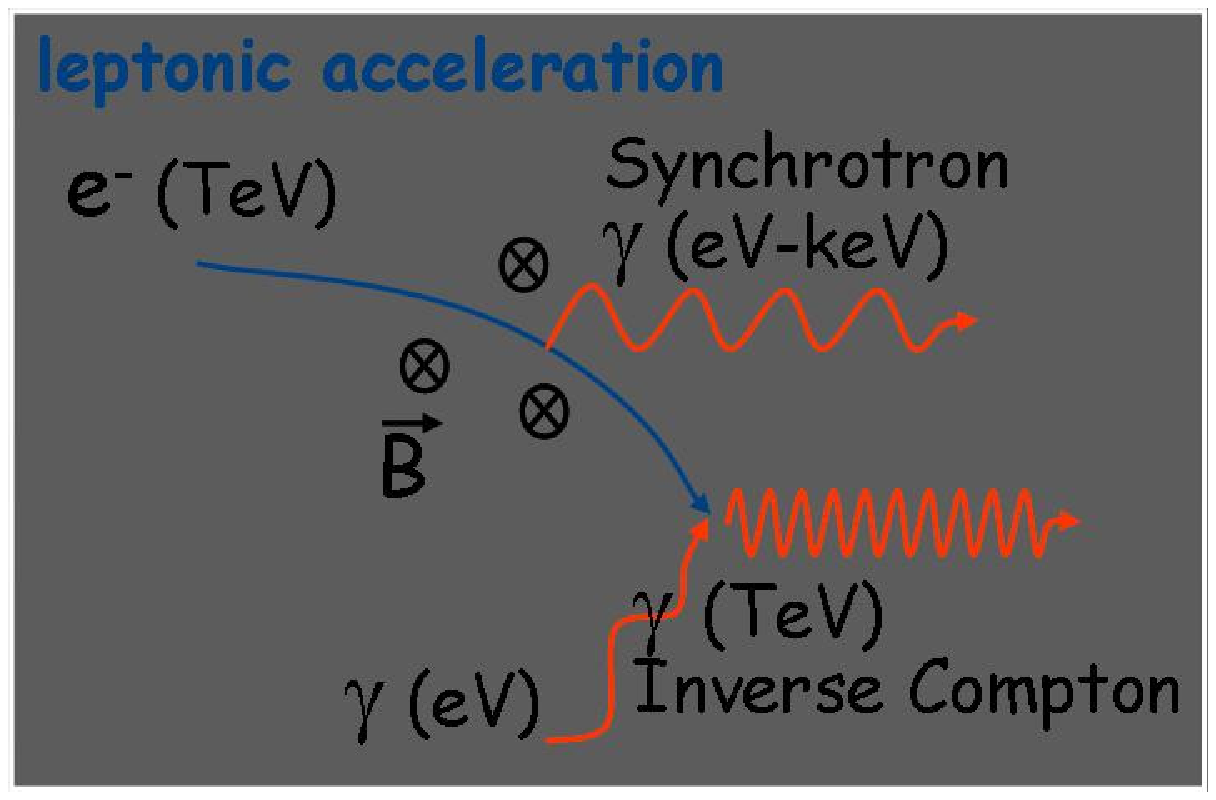} \hfill \includegraphics[width=7cm]{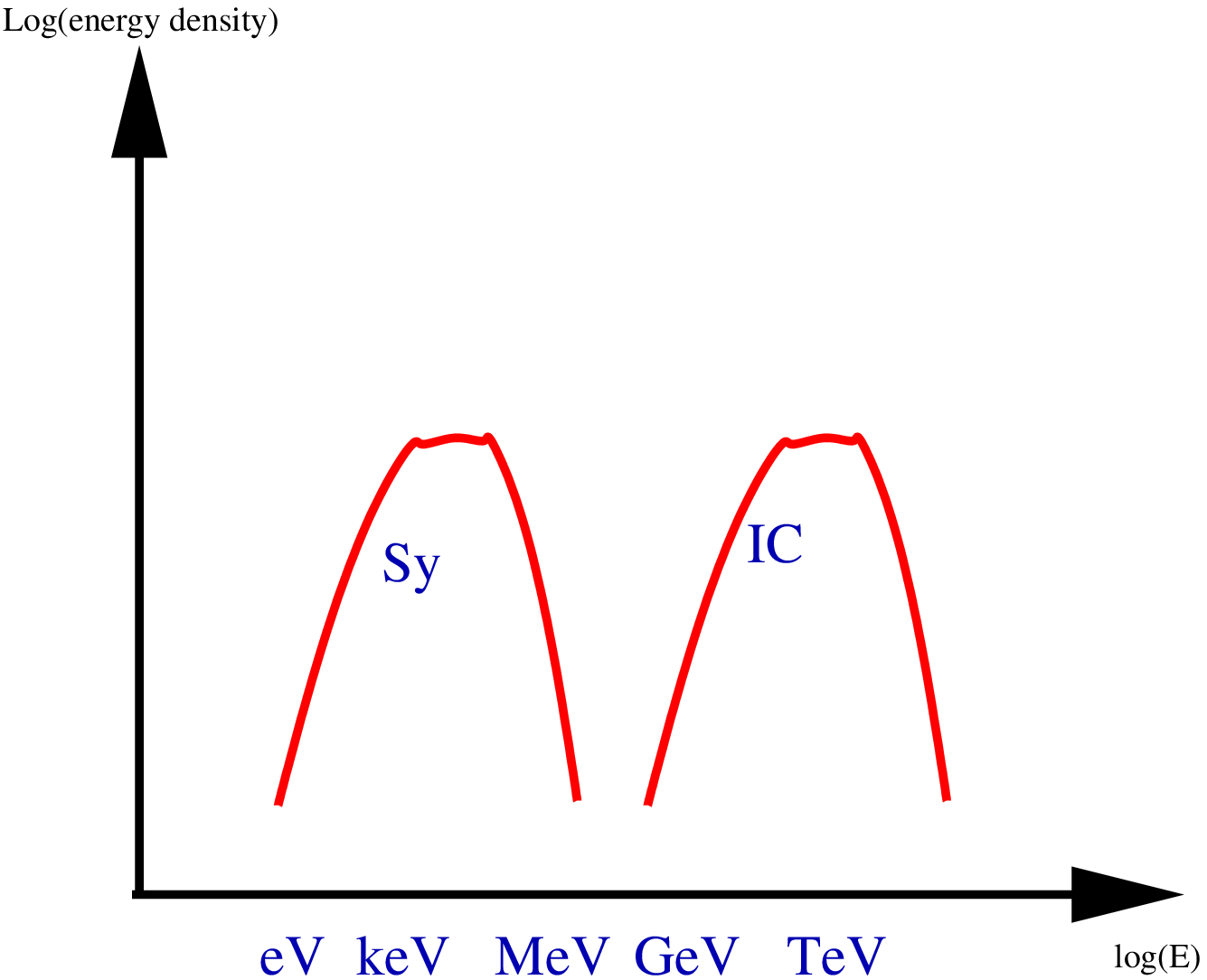}
  \vspace{0.2cm} \includegraphics[width=7cm]{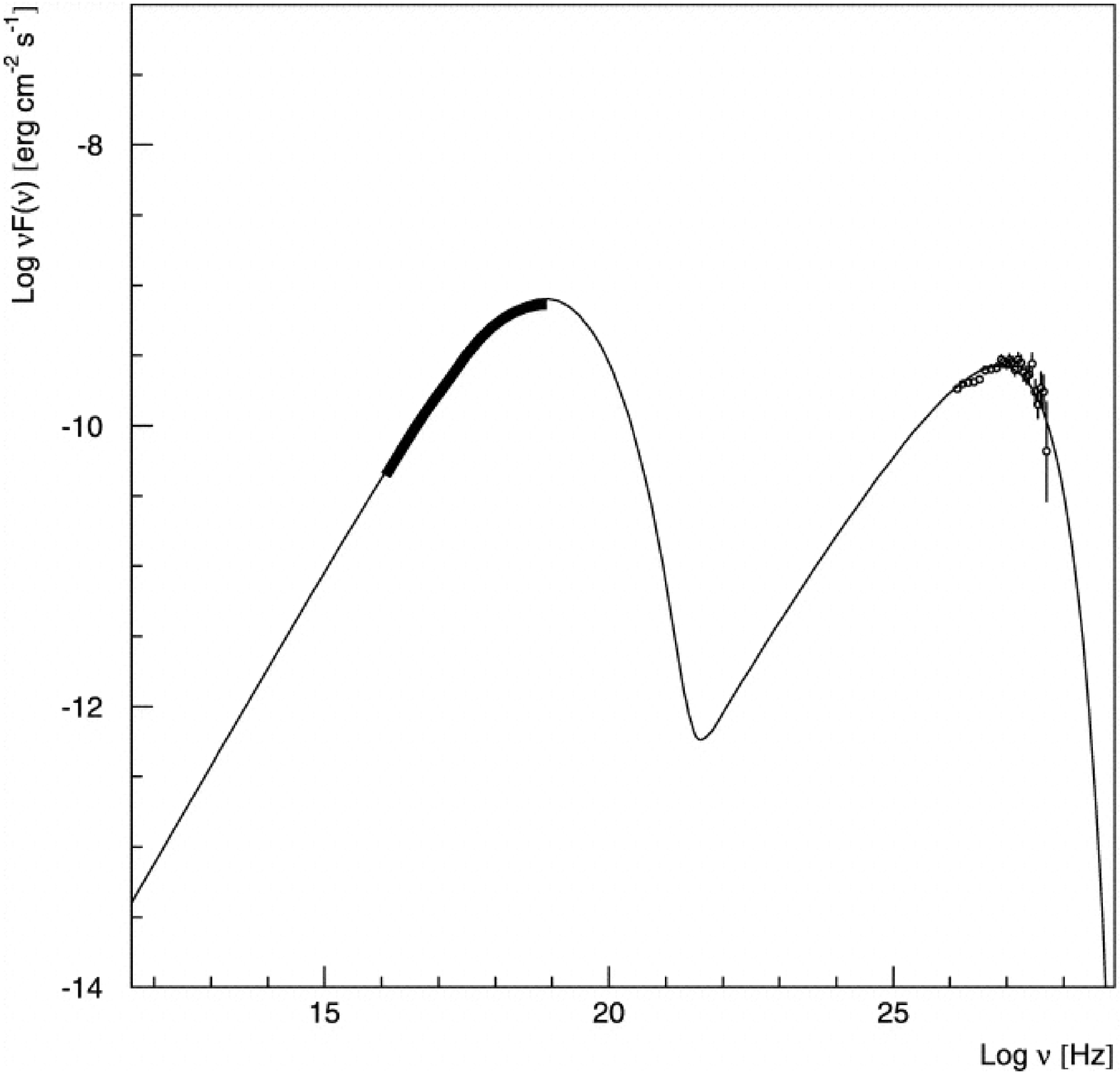}
  \end{center}
\caption{One of the suggested models (leptonic acceleration) for the production of very high energy gamma rays: Synchrotron-self-Compton (SSC), involving synchrotron (Sy) and inverse Compton (IC) scattering for the same electron, provides a mechanism for the production of photons with energies up to several TeV. In the middle figure the typical energy spectrum is indicated, with the characteristic
double peak of the SSC mechanism, one of the synchrotron radiation and the other for the IC scattering. The Inverse Compton spectrum has a peak at TeV energies, as those observed by MAGIC when looking at the AGN Mkn 501 (lower picture, from Konopelko \emph{et al}, ApJ. \textbf{597}, 851 (2005)).}
\label{fig:ssc}
\end{figure}

At present there are three major categories believed responsible for the production of very- and ultra- high-energy Gamma rays:
\begin{itemize}

\item{(i)} Photons from conversion of gravitational energy in Active Galactic Nuclei (AGN) (c.f. fig.~\ref{fig:crproduction}),

\item{(ii)} From self-annihilation of Dark Matter and

\item{(iii)} From decays of exotic massive particles (with masses of order $10^{15}-10^{16}$ GeV/c$^2$), appearing in Grand Unifying Models Beyond the Standard Model (such as string-theory inspired models), in the very Early Universe.

\end{itemize}

In what follows we shall restrict ourselves to (i), which most likely pertains to the production of very high energy Gamma rays observed in AGN such as Mkn 501 observed by the MAGIC Telescope. It is believed today that the centres of galaxies contain massive black holes due to matter collapse~\cite{blackholes}, with typical masses in the range $10^{6}-10^{9}$ solar masses. AGN therefore are celestial systems with very high mass density, and it has long been assumed that they consist of a massive black hole, of say $10^8$ solar masses or more, accreting the gas and dust at the center of the galaxy. The gravitational energy liberated during accretion onto a black hole is 10\% of the rest mass energy of that matter and is the most efficient mass–-energy conversion process known involving normal matter (Gravitational Energy Conversion).
Collapsing matter towards this massive central galactic object releases gravitational energy (Gravitational Energy Conversion) and results in spectacular relativistic material jet emissions.
Since the accreting matter has in general a non-trivial angular momentum, angular momentum conservation is responsible for the matter orbiting the black hole and, through energy dissipation, the formation of a material (flat) accretion disk. This also results in material jets with ultra-relativistic particles outflowing the accretion plane (fig.~\ref{fig:crproduction}).

In addition, since the black holes at the centres of the AGN are probably rotating (as a result
of having a non-zero angular momentum (Kerr type)~\cite{blackholes}, due to angular momentum conservation in the formation (collapse) process), one might also speculate that part of the
relativistic jet might be due to the so-called Penrose process, which allows the extraction of energy from a rotating black hole~\cite{penrose}. The extraction of energy from a rotating black hole is made possible by the existence of a region of the Kerr spacetime called the ergoregion, in which a particle is necessarily propelled in locomotive concurrence with the rotating spacetime. In the process, a lump of matter enters into the ergoregion of the black hole and splits into two pieces, the momentum of which can be arranged so that one piece escapes to infinity, whilst the other falls past the outer event horizon into the hole (a rotating black hole has two event horizons, an outer and an inner one). The escaping piece of matter can possibly have greater mass-energy than the original infalling piece of matter. In summary, the process results in a decrease in the angular momentum of the black hole, leading to a transference of energy, whereby the momentum lost is converted to energy extracted. The process obeys the laws of black hole mechanics. A consequence of these laws is that if the process is performed repeatedly, the black hole can eventually lose all of its angular momentum, becoming rotationally stationary.

The particles in the relativistic jet undergo acceleration, but currently
the pertinent mechanism is a matter of debate and active research. Most likely it depends on the source.
In general there are two generic ways of cosmic acceleration.
\bigskip
\begin{center}

\emph{Mechanisms for Cosmic Acceleration}

\end{center}

\bigskip

\emph{Leptonic Acceleration:} Among the particles in the jet are charged electrons, whose paths are curved as a result of the existing magnetic fields in the galactic regions, which accelerate the electrons. The curved path of a charged objects implies, as well known, synchrotron radiation, as a result of energy conservation. Moreover, in AGN's like Mkn 501, the same (high-energy) electron can also undergo inverse-Compton scattering with low-energy photons (with energies of order eV, e.g. photons of the cosmic microwave background radiation that populate the Universe as remnants of the Big-Bang). The terminology inverse-Compton (IC) scattering refers here to the fact that, contrary to the conventional Compton photon-electron scattering, here it is the electron which is the
high energy particle, and whose loss of energy is converted to outgoing radiation (c.f. fig.~\ref{fig:ssc}). This IC outgoing radiation can have very high energies. In fact, the Compton spectrum peak can be at several TeV energies, as observed for the AGN Mkn 501. This combined process, whereby the \emph{same electron} that is responsible for \emph{synchrotron} radiation in AGN also undergoes \emph{IC scattering} to produce high energy photons is known as Synchrotron-self-Compton (SSC) mechanism~\cite{ssc}, and is believed~\cite{deangelis}
-with some variations to be discussed below - that is responsible for the production of the very high energy photons observed in the AGN Mkn 501 (c.f. fig.~\ref{fig:ssc}).
This is the so-called Leptonic acceleration mechanism, to distinguish it from a different type of acceleration, the \emph{hadronic} one, to be discussed below, and which is believed by many as being the main mechanism for extragalactic ultra-high energy cosmic rays. Notably, AGNs are also extragalactic objects and hence hadronic acceleration mechanism may be relevant for the production of very high energy Gamma rays, as alternative scenarios to (or co-existing with ) SSC-leptonic acceleration mechanism described above.

\begin{figure}[ht]
\begin{center}
  \includegraphics[width=7cm, angle=-90]{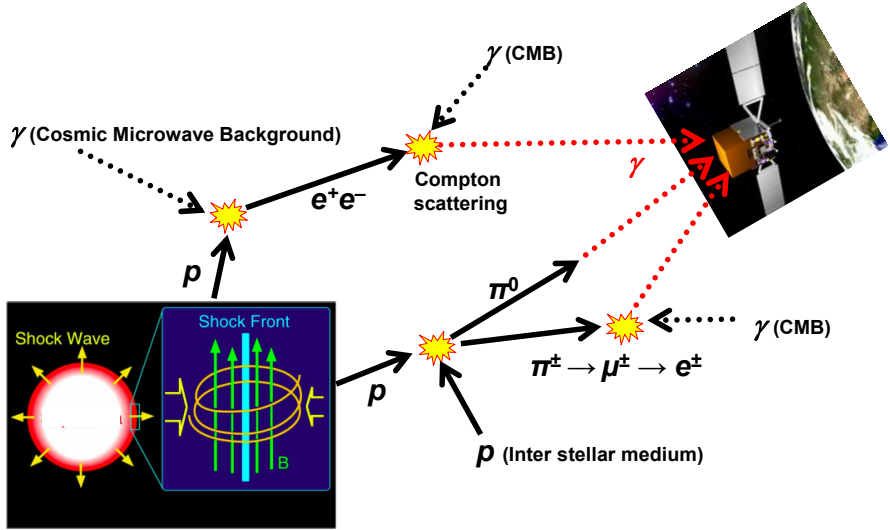}
\end{center}
\caption{A hadronic acceleration model for the production of very high energy gamma rays from extragalactic sources. High-energy Protons which have been accelerated in AGNs or other sources in the presence of magnetic cosmic fields, can interact with the protons of the interstellar medium to give rise to pions (neutral $\pi^0$ and charged $\pi^{\pm}$). Photons from neutral pion decays($\pi^0 \rightarrow 2 \gamma$) could be detected together with those from coming from charged-pion conversion to muon processes $\pi^{\pm} \rightarrow \mu^{\pm} \rightarrow e^{\pm}$, which eventually yield electrons or positrons that scatter \'a la Inverse Compton with low-energy CMB photons resulting in detectable (on Earth or satellite experiments) high-energy photons (picture taken from H. Tajima SLAC-DOE (USA) Programme review talk (June 7, 2006),
(\texttt{www-conf.slac.stanford.edu/programreview/2006/Talks})
).}
\label{fig:hadronic}
\end{figure}
\emph{Hadronic Acceleration:} A prominent way of producing high energy Gamma Rays of extragalactic origin, is the scattering of very high energy protons (produced in the jet of the AGNs by means of gravitational energy conversion mentioned above (c.f. fig.\ref{fig:crproduction}))
off protons in the \emph{interstellar} medium. Such collisions result in pion production, of which neutral pions decay ($\pi^0 \to 2\gamma$) and give rise to very high energy photons
that are detected directly. The charged pions on the other hand are converted to muons, whose decays produces electrons or positrons ($\pi^{\pm} \rightarrow \mu^{\pm} \rightarrow e^{\pm}$); the scattering of the latter with low-energy photons of the cosmic microwave background radiation (CMB) then results, through (inverse) Compton scattering, in photons being detected on Earth or satellites (c.f. fig.~\ref{fig:hadronic}).
\begin{figure}[ht]
\begin{center}
  \includegraphics[width=7cm, angle=-90]{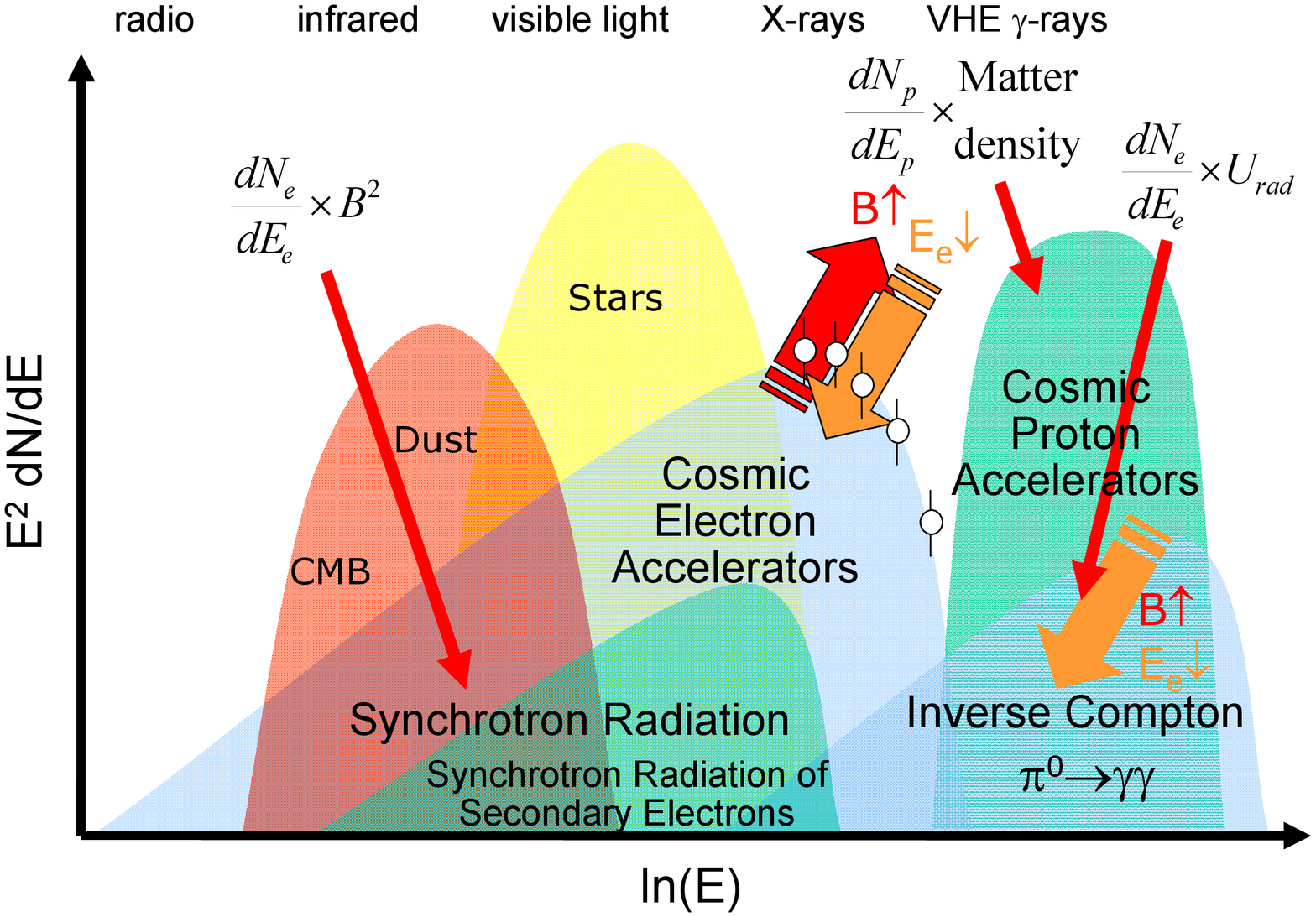}
\end{center}
\caption{Comparing the spectra of very high energy gamma rays in models of leptonic and hadronic acceleration of cosmic particles. In the figure $E$ denotes the photon energy (in some generic units) and $N$ the observed cosmic photon number, while the suffix ``e'' (``p'') denotes quantities pertaining to
electrons (protons) and $B$ denotes the magnetic field. From the differences in the shape and position of these spectra
one can get information on the kind of acceleration that takes place (picture taken from talk of
M. de Naurois, Workshop of HSSHEP, April 2008, Olympia (Greece), (\texttt{http://www.inp.demokritos.gr/conferences/HEP2008-Olympia/})).}
\label{fig:hadron}
\end{figure}

\begin{figure}[ht]
\begin{center}  \includegraphics[width=7cm]{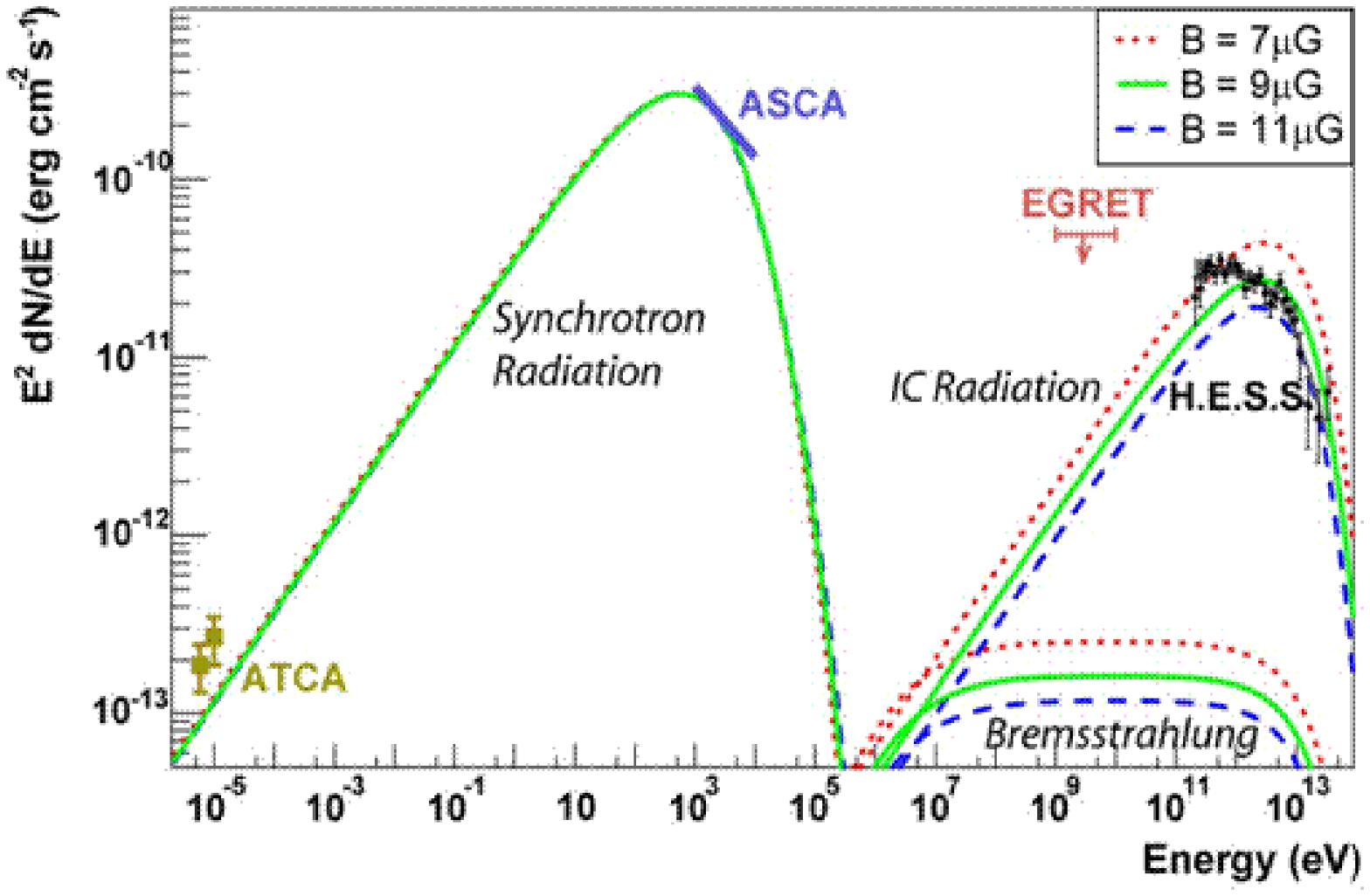} \hfill \includegraphics[width=7cm]{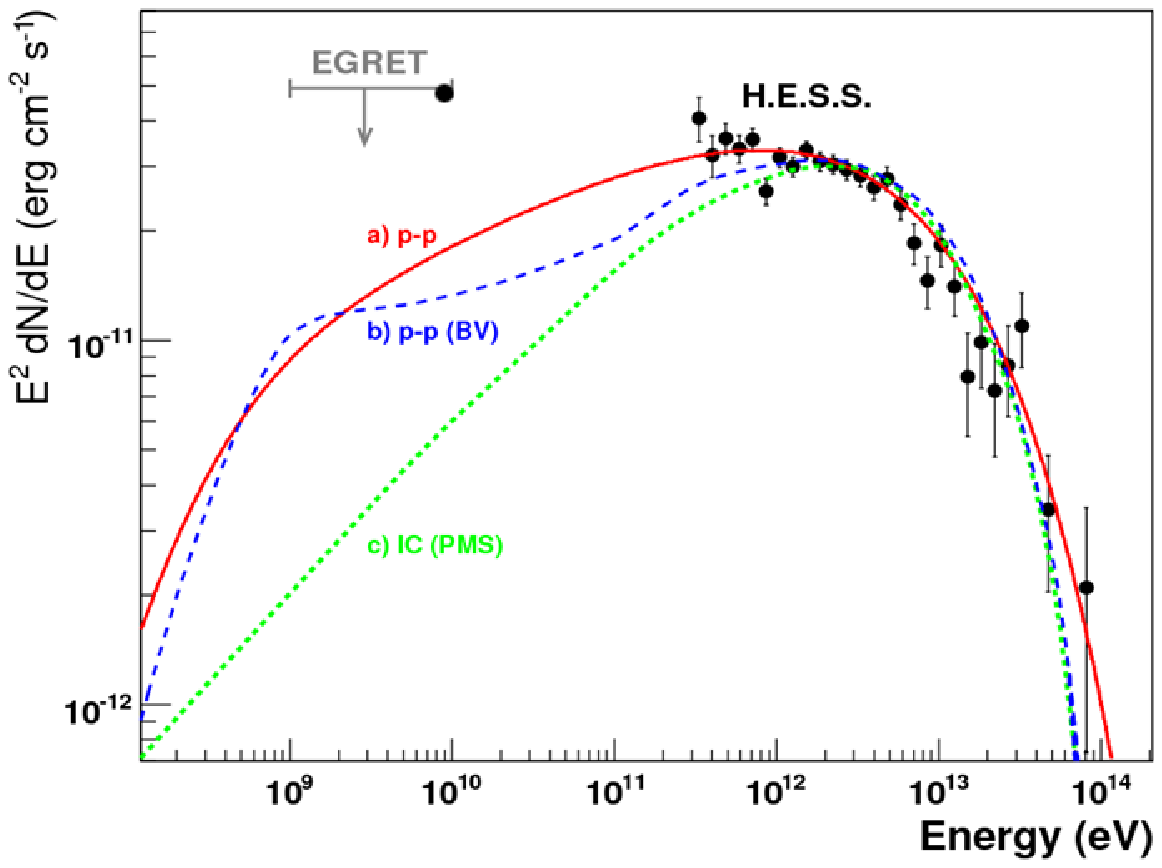}
\end{center}
\caption{ Recent data on Gamma ray spectra. The symbols in the axes have the same meaning as in  fig.~\ref{fig:hadron}. To be sure on the kind of cosmic acceleration taking place (i.e. leptonic or hadronic) one needs data at the lower energy part of the spectrum. Such data can be provided, for instance, by the FERMI (GLAST) satellite, which was recently launched~\cite{glast} (picture taken from talk of
M. de Naurois, Workshop of HSSHEP, April 2008, Olympia (Greece), (\texttt{http://www.inp.demokritos.gr/conferences/HEP2008-Olympia/})).}
\label{fig:hess}
\end{figure}

The energy spectra of hadronic acceleration models are different in shape and location of their peaks in the energy axis from those of leptonic acceleration, as can be seen in fig.~\ref{fig:hadron}, where various photon spectra in the Universe are superimposed for comparison. It is worthy of mentioning that in order to interpret the current high energy gamma ray data using leptonic inverse Compton measurements, one has to assume that the galactic magnetic fields are of low intensity. From measurements of high energy Gamma ray spectra from AGNs or other extragalactic sources, such as Gamma Ray Bursters, we can then soon get sufficient information into the precise way of acceleration of cosmic particles. The current experimental knowledge on
cosmic high-energy gamma ray spectra can be summarised in fig.~\ref{fig:hess}, from which it is clear that we need more measurements in the lower-energy part of the spectrum before conclusions can be drawn on the kind of cosmic acceleration taking place at various celestial sources.

\subsection{The MAGIC delays: Adding to the uncertainties on the VHE Gamma-Ray production mechanisms \label{sec:3magic}}

The observed time delays of order $4 \pm 1$ minutes between the most energetic (TeV) Gamma rays
from AGN Mkn 501 (c.f. fig.~\ref{fig:magic2}), observed by the MAGIC Telescope~\cite{MAGIC,MAGIC2},
lead to further uncertainties in the production mechanism of such photons.

As discussed in \cite{MAGIC}, the conventional model of SSC used to explain the origin of VHE Gamma Rays as being due to an electronic \emph{uniform} acceleration in the AGN jet region, which finds a good application in other AGNs, such as Crab Nebula, fails to account for the observed time delay by MAGIC. The use of the acceleration parameters in the Crab Nebula AGN leads to only millisecond delays of the more energetic photons if applied to the Mkn 501 case.

This prompted speculations that the conventional SSC mechanisms involving
uniform acceleration of the relativistic blob of particles in the jet of the AGN (c.f. fig.~\ref{fig:crproduction}) need to be modified for the Mkn 501 case. In fact, several propositions along this line have been made so far:

\begin{itemize}

\item{(i)} Particles inside the emission region moving with constant
Doppler factor need some time to be accelerated to energies that
enable them to produce $\gamma$ rays with specific high energies in the TeV region
\cite{MAGIC}.

\item{(ii)} The $\gamma$-ray emission has been captured in the initial
phase of the acceleration of the relativistic material blob in the jet of the AGN (c.f.~fig.~\ref{fig:crproduction}), which at
any point in time radiates up to highest $\gamma$-ray energies
possible \cite{bwhdgs}.

\item{(iii)} A version of the SSC scenario (termed \emph{one-zone SSC model}), which invokes a brief episode of increased particle injection at low energies \cite{mas08}. Subsequently,
    the particles are accelerated to high energies, which thus accounts for the observed delays, but they also emit synchrotron and SSC, thereby loosing energy. As in scenario (ii) above, also according to this model the MAGIC observations have caught the relativistic electrons in the jet of the AGN at their acceleration phase.

\end{itemize}

To the above I would also like to add the possibilities that some hadronic mechanisms might also be in operation here, which seems not to have been discussed by the community so far.
It therefore becomes clear from the above brief discussion that the situation concerning the delayed production of VHE Gamma Rays from the AGN Mkn 501 is far from being resolved by means of conventional (astro)physics at the source.

This brings us to the main topic of this article, which is a possible (albeit speculative at this stage) link of the MAGIC observation with more fundamental physics associated with the very structure of space time on which the \emph{propagation} of the VHE Gamma Rays takes place.

\subsection{Quantum-Gravity Space-Time Foam and the MAGIC delays: wild speculation or realistic scenarios? \label{sec:3magicqg}}

In \cite{MAGIC2} it has been observed that, as a result of the ability of the experiment to measure individual (within the accuracy of the observations of course) photons from Mkn 501, of various energies, it should be possible to reconstruct the peak of the flare of July 9th 2005 using
dispersion relations of these individual photons during their journey from emission till observation. In fact, we went one step further and assumed sub-luminal modified dispersion relations of the type expected~\cite{aemn,horizons,mitsou} to be encountered in a model of quantum-gravity (QG) induced space-time foam~\cite{wheeler} coming from string theory~\cite{emnnewuncert}. As we shall discuss below, when we describe in detail this model, the sub-luminality of the QG-induced refractive index in such models is guaranteed by the very nature of string theory, which respects the cornerstone of special relativity that the speed of light in vacuo is the maximal material velocity. In this respect the space-time foam in such theories leads to the \emph{absence of birefringence}, in other words the refractive index is the same for
both photon polarizations. This is an important feature, which allows the MAGIC results to be compatible with other stringent limits of Lorentz invariance from other astrophysical sources, as we shall explain below.

In \cite{MAGIC2} we examine two cases of QG-induced modified dispersion for photons, stemming from (\ref{mdr}) upon assuming:

\begin{itemize}

\item{Case I:} Photon Refractive index  \emph{suppressed  Linearly} by the QG energy scale $M_{{\rm QG1}}$, i.e. only the coefficient $c_1 > 0$ in the series of Eq.~(\ref{mdr}) is non zero, its positivity being required by the sub-luminal nature assumed for the propagation, as ensured by the string theory underlying model~\cite{aemn,mitsou,emnnewuncert}.

\item{Case II:} Photon Refractive index  \emph{suppressed  Quadratically}  by the QG energy scale $M_{{\rm QG2}}$, i.e. only the coefficient $c_2 > 0$ in the series of Eq.~(\ref{mdr}) is non zero, its positivity again being linked to the sub-luminal nature assumed for the propagation. String theory models with this kind of quadratic suppression also exist in the modern approach to string theory, including representation of our world as a brane (domain wall hyperplane)~\cite{pasipoularides}, but will not be discussed here.
\end{itemize}

The method of reconstructing the peak (``most active part'') of the flare by implementing
modified dispersion relations for individual photons is based on the following
well known fact of classical electrodynamics~\cite{jac}: a pulse of electromagnetic radiation
propagating through a linearly-dispersive medium, as postulated above, becomes
diluted so that its power (the energy per unit time) decreases. The
applicability of classical electrodynamics for estimating the low-energy
behavior induced by space-time foam and
the corresponding pulse-broadening effect have been discussed in the paper by J.~Ellis \emph{et al.} (2000) in ref.~\cite{mitsou}, where we refer the interested reader for further details and explicit examples.

The dilution effects
for the linear or quadratic cases may easily be obtained as described in
\cite{jac} by applying the dispersion laws
\begin{equation}
\omega(k)=k[1-k/(2M_\mathrm{QG1})~,  ~\quad {\rm or} \quad ~
\omega (k)=k[1-k^2/(3M_\mathrm{QG2}^2)]~,
\label{mqndisp}
\end{equation}
where $\omega $ denotes the frequency of the photon, with wave vector $k$.
Any transformation of a signal to reproduce the undispersed signal tends to
recover the original power of the pulse. If the parameter $M_{\rm QGn}~, n=1,2$
is \emph{chosen correctly}, the power of the recovered pulse is \emph{maximized}.

\begin{figure}[ht]
\begin{center}\includegraphics[width=0.4\linewidth]{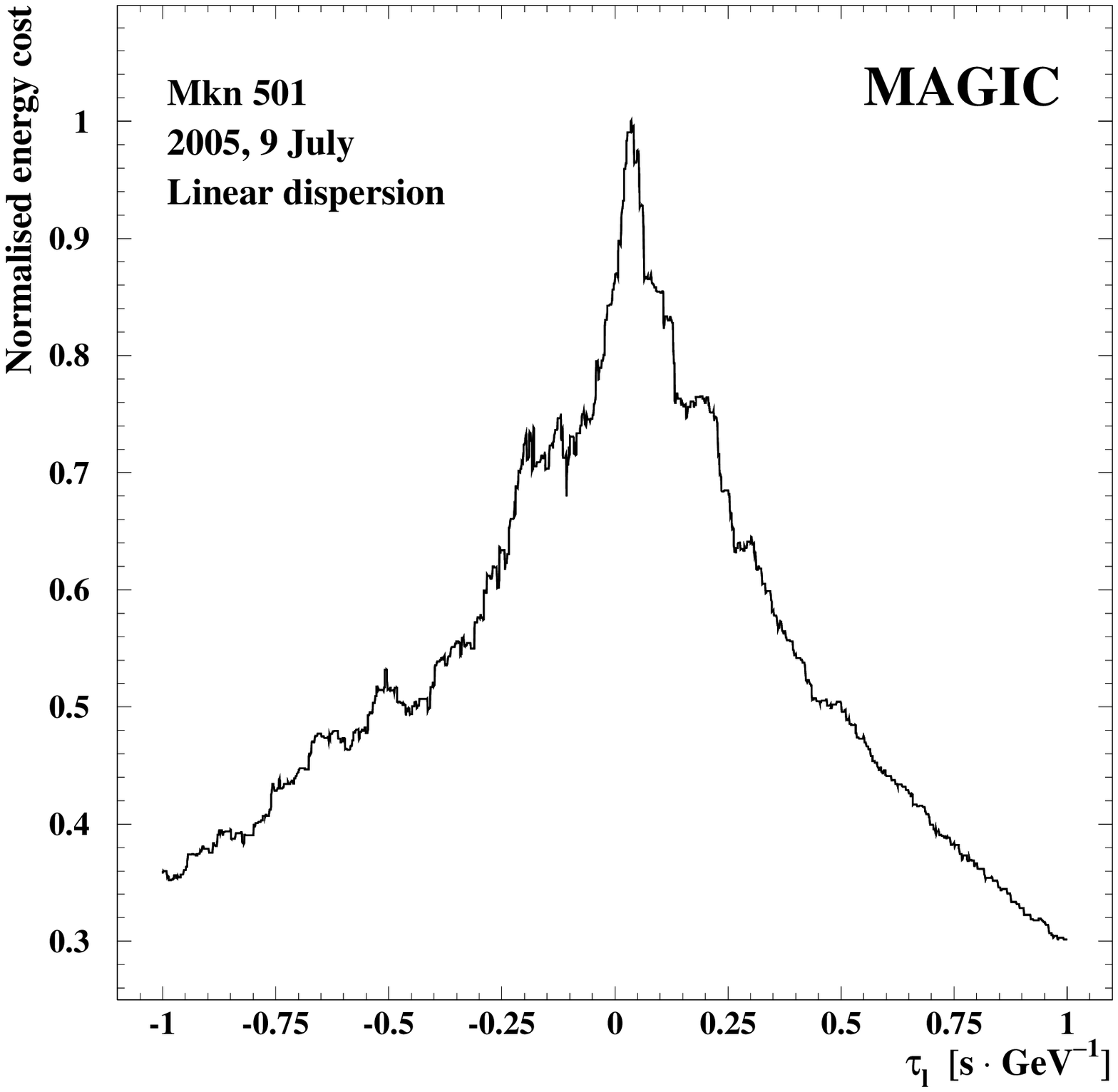}\hfill\includegraphics[width=0.4\linewidth]{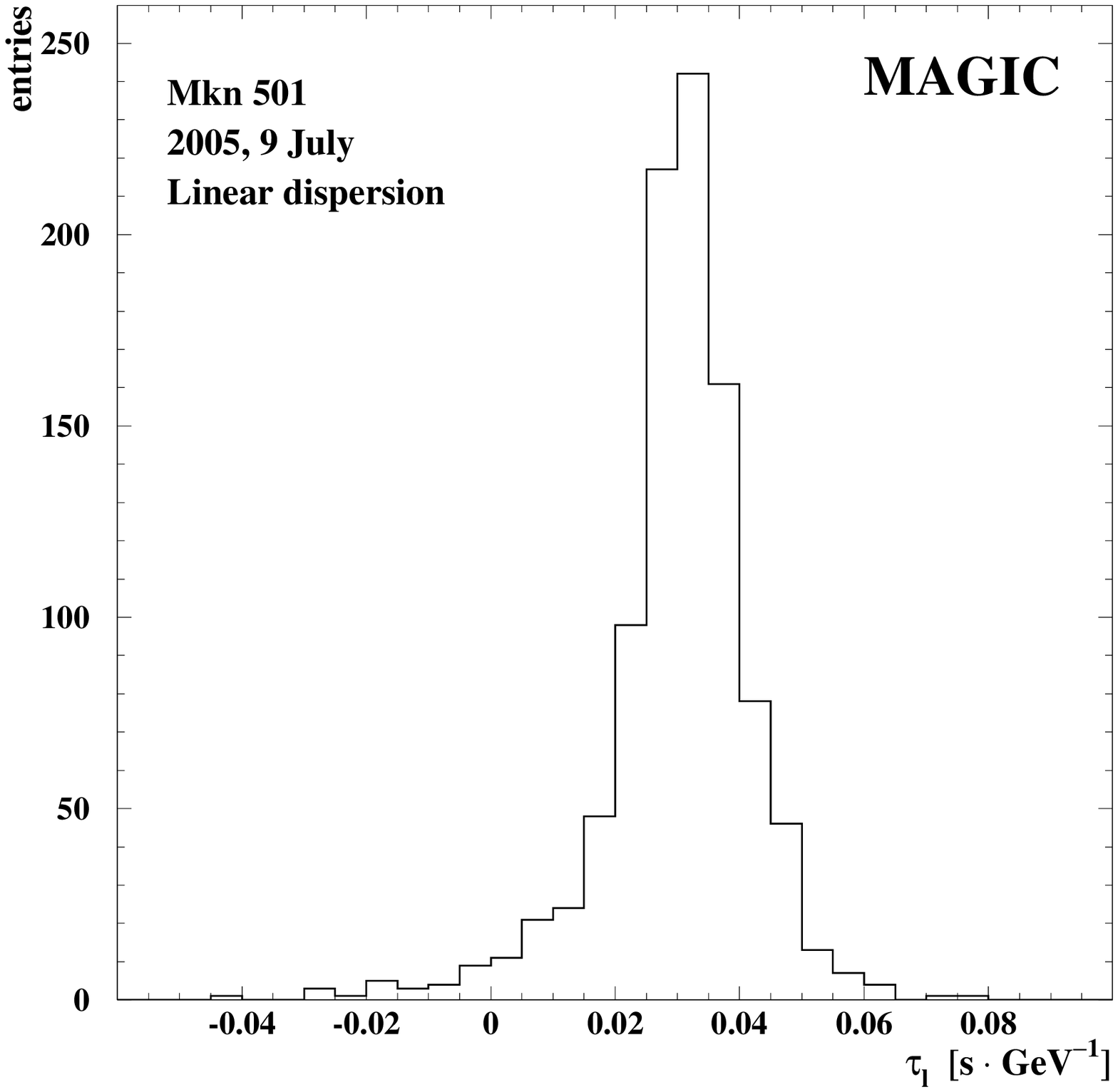}
\end{center}\caption{The \emph{Left figure} shows the Energy Cost Functional (ECF) from one realization~\cite{MAGIC2} of the MAGIC measurements with photon energies smeared by Monte Carlo, for the case of a vacuum refractive index that is linear in the photon energy (Case I). The ECF
exhibits a clear maximum, whose position may be estimated by fitting it with a
Gaussian profile in the peak vicinity. The \emph{Right figure} shows the results of
such fits to the ECFs with $\tau_l$ for the 1000 energy-smeared realizations of
the July~9 flare. From this distribution we derive the value $\tau_l=(0.030\pm
0.012)$~s/GeV, where $M_{\rm QG1}=1.445\times10^{16}\,{\rm s}/\tau_l$, leading
to a lower limit $M_{\rm QG1} > 0.21 \times 10^{18}$~GeV at the 95\%
C.L.}
\label{fig:reconstr}
\end{figure}

This observation has been implemented in the analysis of \cite{MAGIC2}
by appropriately choosing (using statistical-analysis techniques)
a time interval $(t_1;t_2)$ containing the most active part of the flare.
For the record, we mention that this procedure has been applied in \cite{MAGIC} to 1000 Monte Carlo (MC) data samples generated by applying to the measured photon energies the (energy-dependent)
Gaussian measurement errors. The results of the reconstruction of the peak (``most active part'') of the flare using the linear Case I are demonstrated in fig.~\ref{fig:reconstr} for completeness. In a similar way one gets bounds on the quadratic Case II. These results have also been confirmed using different statistical analysis techniques, independent of the ECF. The interested reader is referred to \cite{MAGIC2} for details of the analysis.

Taking into account the \emph{uncertainties in the production mechanism} at the source,
which could also contribute as we have discussed before to the observed delays, we
can only place lower bounds on the \emph{quantum gravity energy scale} from such an analysis.
In fact for the linear and quadratic cases we obtain the lower bounds
(c.f. fig~\ref{fig:reconstr}):
\begin{eqnarray}
M_{\rm QG1} &>& 0.21 \times 10^{18}~ {\rm GeV }~~{\rm at~95\%~Confidence~Level~(C.L.)},~\nonumber \\
M_{\rm QG2} &>& 0.26 \times 10^{11}~ {\rm GeV}~~ {\rm at~95\%~C.L}..
\label{mq12bound}
\end{eqnarray}
It is important to notice that, had the mechanism at the source been understood, and the emission of the different photons been
more or less simultaneous, i.e. by at least two orders of magnitude smaller as compared to the observed delays, then the above lower bounds could be turned into a real measurement of the
quantum gravity scale. It is surprising, therefore, that in such a situation Case I can reproduce the observed delays, provided $M_{{\rm QG1}}$ is of order of the so-called \emph{reduced Planck mass}, which is an energy scale characterising conventional string theory models.

In the analysis of \cite{MAGIC2} it was possible to exclude the possibility that the observed time delay may be due
to a conventional QED plasma refraction effect induced as photons propagate
through the source. From the discussion in sub-section (\ref{sec:ntv}), in
particular eq.~(\ref{highenergy}), it becomes clear that
if the delay would be due to plasma effects at the source region, then
this would induce $$\Delta t = D (\alpha^2
T^2/6k^2) \ln^2(kT/m_e^2)~,$$ where $\alpha$ is the fine-structure constant, $k$
is the photon momentum, $T$ is the plasma temperature, $m_e$ is the mass of
electron, $D$ is the size of the plasma, and we use natural units: $c, \hbar
=1$. Plausible numbers such as $T \sim 10^{-2}$~MeV and $D \sim 10^9$~km
(for a review see \cite{hillas}) yield a negligible effect for $k \sim 1$~TeV, which
are the photon momenta relevant to the MAGIC experiment.
Exclusion of other source effects, such as time evolution in the mean emitted
photon energy, might be possible with the observation of more flares, e.g., of
different AGNs at varying redshifts. Observations of a single flare cannot
distinguish the quantum-gravity scenarios considered here from modified
synchrotron-self-Compton mechanisms.

\subsection{H.E.S.S. and FERMI Observations and Quantum-Gravity Scale Bounds}

However, the above-described
pioneering study demonstrates clearly the potential scientific value of an
analysis of multiple flares from different sources. The most promising
candidate for applying the analyses proposed here is the flare
from the Active Galaxy PKS 2155-304 detected recently by H.E.S.S. (\emph{H}igh \emph{E}nergy \emph{S}tereoscopic \emph{S}ystem) Collaboration~\cite{hess2155}, another
experiment involving arrays of Cherenkov Telescopes in Namibia (Africa). This galaxy lies further than Mk501 at redshift $z = 0.116$ and there is a much higher statistics of photons at energy ranges of a few TeV.

In fact, quite recently, H.E.S.S. collaboration published their measurements~\cite{hessnew} on the arrival time of photons from PKS 2155-304.
However, unlike the MAGIC observations from Mk501 Galaxy, there was no time lag found between higher- and lower-energy photons in this case. These results can thus place only bounds on the quantum gravity scale,
if space-time foam is assumed to affect the photon propagation. The bounds are similar to the MAGIC case (\ref{mq12bound}) above.

The H.E.S.S. result can mean several things, and certainly points towards different source mechanisms for the acceleration of photons between the two galaxies Mkn 501 and PKS 2155-304. However, this second measurement by H.E.S.S. cannot still rule out the possibility that quantum gravity plays a r\^ole in photon propagation. For instance, one cannot exclude the (admittedly remote) possibility that, due to a still unknown source effect, the high-energy photons in the PKS 2155-304 Galaxy are emitted first, in contrast to the Mkn 501 case, in such a way that a sub-luminal vacuum refractive index quantum-gravity effect, of a strength appropriate to produce the delays observed by the MAGIC experiment, ``slowed these photons down'' as compared to the lower-energy ones, so that there are no observable delays in the arrival times between the higher and lower energy photons in the H.E.S.S. experiment. Of course, one cannot exclude the possibility that the conditions of this set of measurements, for some reason, prohibited the detection of an observable time lag between high and low energy photons.

Although the available sample of AGNs is still not large enough for a robust
analysis on bounds of Quantum-Gravity medium effects, nevertheless, one can at least check for consistency between the
available MAGIC and HESS results, and gauge the magnitude of possible intrinsic
fluctuations in the AGN time-lags. Comparing the time-lag measured by MAGIC for
Mkn 501 at redshift $z = 0.034$: $\Delta t/E_\gamma = 0.030 \pm 0.012$~s/GeV, with that
measured for PKS 2155-304 at $z = 0.116$: $\Delta t/E_\gamma = 0.030 \pm 0.027$~s/GeV,
we see that they are compatible with a common, energy-dependent {\it intrinsic} time-lag
at the source.
On the other hand, they are also compatible with a universal redshift- and energy-dependent
{\it propagation} effect:
\begin{equation}
\Delta t/E_\gamma = (0.43 \pm 0.19) \times K(z) {\rm s/GeV} ,~\quad
K(z) \equiv   \int_0^z \frac{(1 + z)dz}{\sqrt{\Omega_\Lambda + \Omega_m (1 + z)^3}} ,
\label{bestfit}
\end{equation}
assuming an expanding Universe within the framework of the standard Cosmological-constant-Cold-Dark-Matter ($\Lambda$CDM) model.
The best fit of the MAGIC and HESS data based on (\ref{bestfit}) leads to the following result for the
Quantum Gravity scale, assuming that it is the dominant cause of the delay:
$M_{QG1} = (0.98^{+0.77}_{-0.30}) \times 10^{18}$~GeV.

With measurements from only a few available flares from AGN it is,
therefore, not possible to disentangle with any certainty source from propagation effects.
For this we need statistically significant
populations of available data.
Unfortunately the occurrence of fast flares in AGNs is currently unpredictable,
and since no correlation has yet been established with observations in other
energy bands that could be used as a trigger signal, only serendipitous
detections are currently possible.
It seems unlikely that the relatively rare
and unpredictable sharp energetic flares produced
only occasionally by AGNs, which have a relatively restricted redshift range and hence a
small lever arm, will soon be able to provide the desired discrimination.

On the other hand, Gamma Ray Bursts (GRBs) are observed at a relatively high rate, about one a day, and
generally have considerably larger redshifts.
The advent of the FERMI (n{\' e}e GLAST) Telescope with its large acceptance offers the
possibility of achieving the required sensitivity. Indeed, the FERMI Collaboration has
already made a preliminary report of GeV-range $\gamma$ rays from the GRB 080916c~\cite{grbglast}.
In this GRB, there is a 4.5-second time-lag between the onsets
of high- ($> 100$~MeV) and low-energy ($< 100$~KeV)
emissions. Moreover, the highest-energy photon GRB 080916c measured by the FERMI $\gamma$-ray telescope had an energy $E = 13.2^{+0.70}_{-1.54}$~GeV, and was
detected $\Delta t = 16.5$~s after the start of the burst.
Spectroscopic information has been used by the GROND Collaboration~\cite{GROND} to estimate
the redshift of GRB 080916c as $z = 4.2 \pm 0.3$~\cite{grbglast}. Assuming
this value of the redshift, the best fit (\ref{bestfit}) would correspond to
a time-lag
\begin{equation}
\Delta t = 25 \pm 11~{\rm s}
\end{equation}
for a 13~GeV photon from GRB 080916c. As discussed in \cite{emnnewuncert,grbglast}, such time delays can fit excellently within the above-mentioned QG scenario for a subluminal refractive index for photons, with the following lower bound for the QG scale
$M_{QG1} > 1.50 \pm 0.20 \times 10^{18}~{\rm GeV}$, where the inequality is due to the ignorance of the source mechanism.
This bound is consistent with the MAGIC
and HESS results stated previously. The reader should also bear in mind that
the 4.5-second time-lag
observed for $\sim 100$~MeV photons could not be explained by a propagation
effect that depends linearly on the energy~\cite{grbglast}. Because of ignorance of the source
mechanism, the preliminary analysis of these data by the FERMI Collaboration~\cite{grbglast} quoted a lower bound
$M_{QG1} > 1.50 \pm 0.20 \times 10^{18}~{\rm GeV}$, which is consistent with the MAGIC
and HESS results stated previously. It is clear therefore that the FERMI Telescope has already demonstrated the sensitivity to probe a possible linearly
energy-dependent propagation effect at the level reached by the available AGN data,
and it is appropriate and possibly helpful to consider how such an effect could be probed
in the future.

The analysis of the second reference in \cite{emnnewuncert} has also demonstrated that although the three above-mentioned sets of data, from MAGIC, FERMI and HESS Collaborations, can be explained simultaneously by a linear in energy vacuum refractive index, suppressed by a single power of the quantum gravity scale, this is not the case for a refractive index scaling quadratically with the photon energy, for instance of the type encountered in some brane models with asymmetric warp factors, as in ref.~\cite{pasipoularides}. Indeed, on assuming that the quadratic refractive index is the sole cause of the observed delays in the arrival of high- vs. low -energy photons
in \emph{both} the MAGIC and FERMI cases, this would imply a time delay of order $0.24 \pm 0.16$ s for the most energetic photon (13.22 GeV) of the GRB 080916c, for a quantum gravity scale that saturates $M_{\rm QG2}$ in (\ref{mq12bound}).
This is two orders of magnitude smaller than the measured time-lag (16.5 s) by the FERMI Collaboration~\cite{grbglast}.

Prompted by the MAGIC (and subsequently FERMI) results, in particular their consistency with a
non-trivial linear in energy vacuum refractive index, the authors of \cite{emnnewuncert} attempted to discuss space-time foam models within the framework of string theory and in particular its modern extension, involving
solitonic domain-wall defects (D(irichlet) branes. We review such a model (or better a class of such models) in the next section. Before embarking on such a task, however, we consider it as essential to compare the bounds
(\ref{mq12bound}) on QG-induced anomalous photon dispersion (\ref{mqndisp})
 obtained from the MAGIC observations with similar bounds on Lorentz Violation from other astrophysical processes. This will be important for the viability of the string model as a possible explanation of the
 MAGIC ``anomaly''. Indeed, it is important to notice that, if the results of MAGIC (and FERMI) collaborations are attributed as being predominantly due to the ``\emph{medium}'' of Quantum Gravity, which leads to anomalous photon dispersion,
 then it is imperative that this medium is transparent to electrons, and it does not imply birefringnece effects, since the latter would lead to much more stringent bounds on the anomalous terms, incompatible with the bounds (\ref{mq12bound}) from photon measurements.

\subsection{MAGIC and FERMI time-delays versus other (astrophysical) constraints on quantum-gravity foam \label{sec:biref}}

The sensitivity of the MAGIC (and FERMI) observations to Planck scale physics (\ref{mq12bound}), at least
for linearly suppressed modified dispersion relations, calls for an immediate comparison with other sensitive probes of non-trivial optical properties of QG medium.

Indeed, from the analysis of \cite{MAGIC2}, there was no microscopic model dependence of
the induced modifications of the photon dispersion relations, other than the sub-luminal nature of the induced refractive index and the associated absence of birefringence, that is the independence
of the refractive index on the photon polarization. The latter feature avoids the otherwise very stringent constraints on the photon dispersion relation imposed by astrophysical observations, as we now come to discuss.

We shall be very brief in our description of the complementary astrophysical tests on Lorentz invariance and quantum-gravity modified dispersion relations, to avoid large diversion from our main point of this review article which is string theory.

There are three major classes of complementary astrophysical constraints, to be considered in any attempt to interpret the MAGIC, FERMI or more general $\gamma$-ray Astrophysics results in terms of quantum-gravity induced anomalies in photon dispersion.

\begin{itemize}

\item{\emph{Birefringence and strong constraints on QG-induced photon dispersion:}}

In certain models of quantum gravity, with modified dispersion relations, for instance the so-called loop-quantum gravity~\cite{gambini}, the ground state breaks reflexion symmetry (parity) and this is one of the pre-requisites for a dependence of the induced refractive index on the photon polarization, \emph{i.e}. birefringence. We remind the reader that in birefringent materials this is caused precisely by the existence of some kind of anisotropies in the material. The velocities of the two photon polarizations (denoted by $\pm$) in such QG models may be parametrised by:
\begin{equation}
v_{\pm} = c\left(1 \pm \xi (\hbar \omega /M_P )^n \right)
\label{birefrrel}
\end{equation}
where $M_P = 1.22 \times 10^{19}$ GeV is the Planck energy scale, and $\xi$ is a parameter
following from the underlying theoretical model, which is related with the modifications
of the pertinent dispersion relations for photons. The order of suppression
of these effects is described by $n$ which in the models of \cite{gambini} assumed the value $n=1$, but in general one could have higher order suppression, as we have discussed in (\ref{mdr}).

 Vacuum QG birefringence should have showed up in optical measurements from remote astrophysical sources, in particular Gamma Ray Bursters (GRB). The latter are cosmic explosions of titanic proportions, due to collapsing massive stars at distant parts of the universe.

Ultraviolet (UV) radiation measurements from distant galaxies~\cite{uv} and UV/optical polarization measurements of light from  Gamma Ray Bursters~\cite{grb} rule out
birefringence unless it is induced at a scale (way) beyond the Planck mass (for linear models
the lower bound on the QG scale in such models can exceed the Planck scale ($\sim 10^{19}$ GeV) by as much as \emph{seven orders of magnitude}). Indeed,  in terms of the parameter $\eta$ introduced above (c.f. (\ref{birefrrel})), for the case $n=1$ of \cite{gambini}, one finds from
 optical polarization observations that the absence of detectable birefringence effects imply the upper bound  $|\xi| < 2 \times 10^{-7}$, which is incompatible with the MAGIC observed delays,
 saturating from below the bounds (\ref{mq12bound}).

 At this point, we wish to mention that, using recent polarimetric observations of the Crab Nebula in the hard X-ray band by INTEGRAL~\cite{integral}, the authors of \cite{xray} have demonstrated  that the absence of vacuum birefringence effects constrains linearly suppressed Lorentz violation in quantum electrodynamics to the level $|\xi | < 6 \times 10^{-10}$ at 95\% C.L., thereby tightening by about three orders of magnitude the above-mentioned constraint.

\item{\emph{Synchrotron Radiation and further stringent constraints for electronic QG-induced anomalous dispersion in vacuo.}}

\begin{figure}[ht]
\begin{center}
\includegraphics[width=0.2\linewidth]{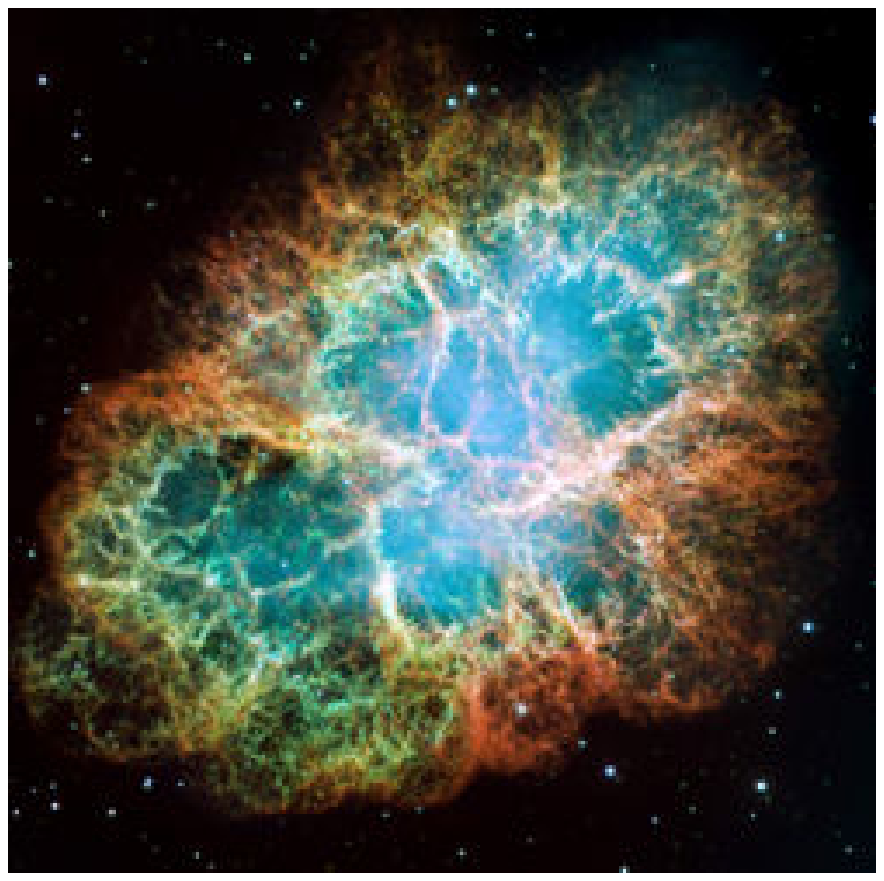}\hfill \includegraphics[width=0.2\linewidth]{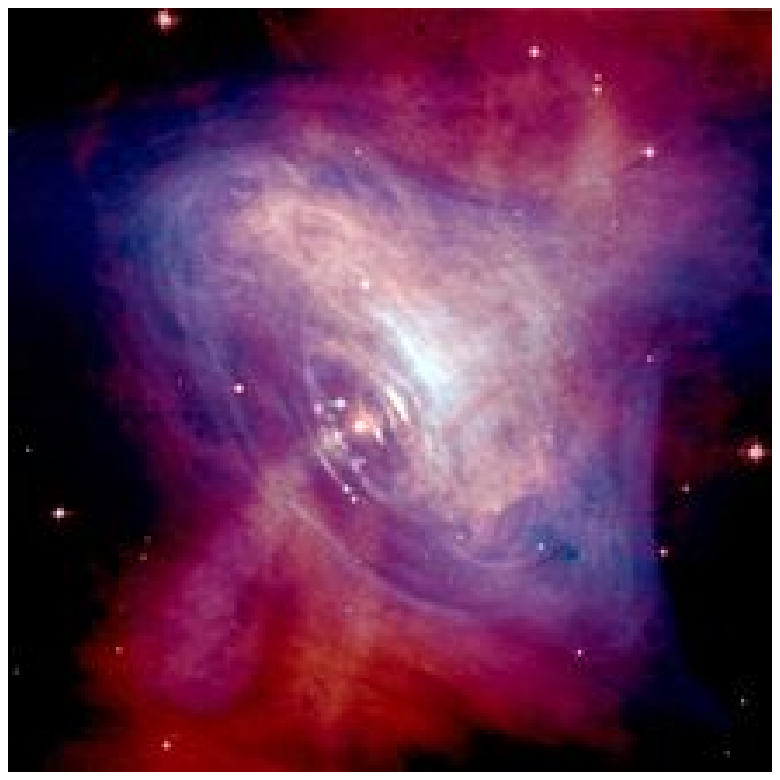}\hfill
\includegraphics[width=0.4\linewidth]{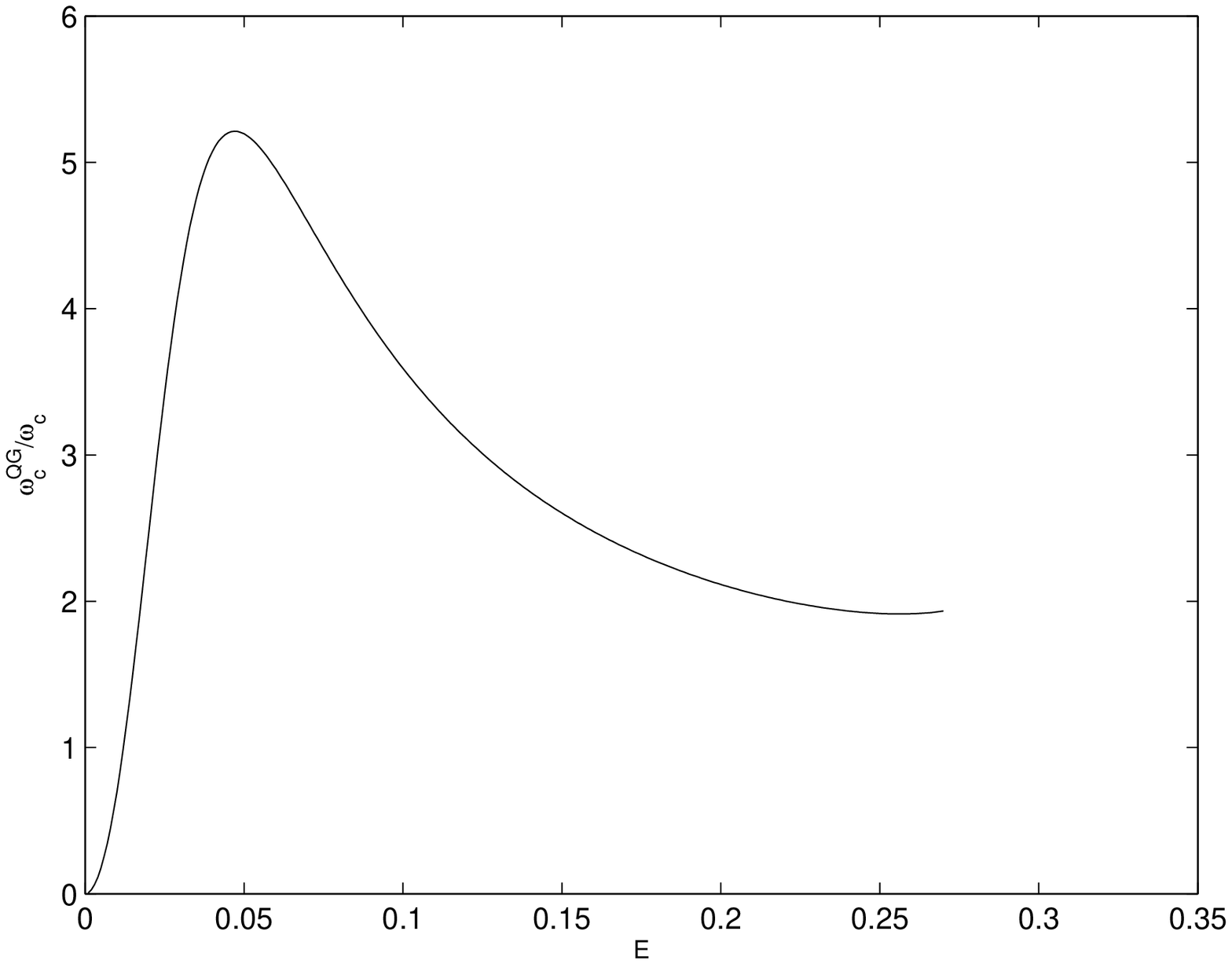}
\end{center}
\caption{The Crab Nebula (left image) is a supernova remnant, with a rotating neutron star (the Crab Pulsar) at its centre (middle image). Observations of synchrotron radiation (right image, radiation spectrum (arbitrary units)) from such celestial objects places very stringent constraints on quantum gravity models with anomalous dispersion relations for electrons (images taken from NASA, \texttt{http://www.nasa.gov/} (Crab Nebula) and wikipedia, \texttt{http://en.wikipedia.org/wiki/CrabNebula} (Crab Pulsar)) and \cite{ems} (synchrotron spectrum).}
\label{fig:crab}
\end{figure}
Another important experimental constraint on models with QG-induced anomalous dispersion relations
comes from observations of synchrotron radiation from distant galaxies~\cite{crab,ems,crab2}, such as Crab Nebula (c.f. fig.~\ref{fig:crab}). As mentioned previously, the magnetic fields at the core regions of galaxies curve the paths of (and thus accelerate) charged particles, in particularly electrons (which are stable and therefore appropriate for astrophysical observations), and thus, on account of energy conservation this results in synchrotron radiation.

In standard electrodynamics~\cite{jac}, electrons in an
external magnetic field ${H}$, follow helical orbits
transverse to the direction of ${H}$. The so-accelerated electrons in a magnetic field
emit synchrotron radiation with a spectrum that cuts off sharply at a
frequency $\omega_c$ (c.f. fig.~\ref{fig:crab}, right panel):
\begin{equation}
\omega^{LI}_c = \frac{3}{2}\frac{eH}{m_0}\frac{1}{1 - \beta^2},
\label{sync}
\end{equation}
where $e$ is the electron charge, $m_0$ its mass, and $\beta_\perp \equiv v_\perp $ is the component of the velocity of the electron perpendicular to the direction of the
magnetic field.
The superfix $LI$ in (\ref{sync})
stresses that this formula is based on a LI approach, in which one
calculates the electron trajectory in a given magnetic field $H$ and the
radiation produced by a given current, using the relativistic relation
between energy and velocity.

All these assumptions are
be affected by violations of Lorentz symmetry, such as those encountered in quantum-gravity space-time foam models, leading to modified dispersion relations for
photons and electrons of the form:
\begin{eqnarray}
 \omega^2(k)&=& k^2 + \xi_\gamma \frac{k^{2 + \alpha}}{M_P^\alpha},
 \label{eq:pdr}\\
 E^2(p)&=& m_0^2+ p^2 + \xi_e \frac{p^{2 + \alpha}}{M_P^\alpha},
 \label{eq:mdr}
\end{eqnarray}
for photons and electrons, respectively, where $\omega$ and $k$ are the
photon frequency and wave number, and $E$ and $p$ are the electron energy
and momentum, with $m_0$ the electron (rest) mass. In the spirit of the MAGIC observation analysis above, we assume here linear ($\alpha =1$) or quadratic QG ($\alpha = 2$)
effects, characterized by parameters $\xi_\gamma$ and $\xi_e$, extracting
the Planck mass scale $M_{P}=1.22\times 10^{19}$ GeV. In fact one can do the analysis~\cite{ems} for a general $\alpha$ (single power) and attempt to extract limits on this parameter by matching with observations.

A detailed analysis~\cite{crab,ems}, including the modifications in the electron's trajectories due to space-time foam~\cite{ems}, yields:
\begin{equation}
\omega_c^{QG} \propto \omega_c^{LI} \frac{1}{
(1 + \sqrt{2 -1/{\cal \eta}^2})^{1/2}
\left(\frac{m_0^2}{E^2} + (\alpha + 1)\left(\frac{E}{{\cal M}}\right)^\alpha
\right)}~, \quad {\cal M} \equiv M_P/|\xi_e|~, \quad \eta \equiv
1 - (E/{\cal M})^\alpha~,
\label{modqg}
\end{equation}
where $\omega_c^{LI}$ is given in (\ref{sync}) and the superscript ``QG'' indicates that the QG-modified dispersion relations (\ref{eq:mdr})
are used. This function is plotted schematically (for $\alpha =1$)
in fig.~\ref{fig:crab} (right panel).

In \cite{crab,ems},
the above QG-modified dispersion relations have been tested using observations
from Crab Nebula.
It should be emphasized that the
estimate of the end-point energy of the Crab synchrotron spectrum
and of the magnetic field used above are indirect values based on
the predictions of the Synchrotron Self-Compton (SSC) model
of very-high-energy emission from Crab Nebula~\cite{reports}. In \cite{ems}
the choice of parameters used was the one that gives good agreement
between the experimental data on high-energy emission and the
predictions of the SSC model \cite{reports,hillas}.
Estimating the magnetic
field of Crab Nebula in the region
$160 \times 10^{-6}~{\rm Gauss} ~ < ~ H ~ < ~ 260 \times 10^{-6} ~{\rm Gauss}$,
and requiring $|\xi_e| \le 1$ (which thus sets the quantum gravity scale as at least $M_P$)
one obtains
the following bounds on the exponent $\alpha $ of the dispersion relations (\ref{eq:mdr})~\cite{ems}:
\begin{equation}
\alpha \ge \alpha_c~: \qquad 1.72 < \alpha_c < 1.74
\label{alphabounds}
\end{equation}
These results
imply already a sensitivity
to quadratic QG corrections with Planck mass suppression $M_P$.

However, for photons there are no strong constraints on $\xi_\gamma$ coming from synchrotron radiation studies, unless in cases where QG models entail  birefringence~\cite{crab2}, where, as we discussed above, strong constraints on the photon dispersion are expected at any rate from optical measurements on GRBs.
In this sense, the result (\ref{alphabounds}) \emph{excludes the possibility} that the MAGIC observations
leading to a four-minute delay of the most energetic photons are due to a quantum foam that acts \emph{universally} among photons and electrons. However, the synchrotron radiation measurements cannot exclude anomalous photon dispersion with linear Planck-mass suppression, leading to a saturation of the lower bound (\ref{mq12bound}), in models where the foam is transparent to electrons, as in the string case to be discussed in the next section.

\item{\emph{Strong constraints from Ultra-high-energy Cosmic photon annihilation}}

Further strong constraints on generic modified dispersion relations for photons, like the ones used in the aforementioned QG-interpretation of the MAGIC results~\cite{MAGIC2}, comes from processes of scattering of ultra-high-energy photons, with energies above $10^{19}$ eV off
very-low energy cosmic photons, such as the ones of the cosmic microwave background (CMB) radiation that populates the Universe today, as a remnant from the Big-Bang epoch.
In \cite{sigl} it has been argued that the non-observation of such ultra-high energy (UHE) photons
places very strong constraints on the parameters governing the modification of the photon dispersion relations, that are several order of magnitude smaller than the values required to reproduce the MAGIC time delays, should the effect be attributed predominantly to photon propagation in a QG dispersive medium.

The main argument relies on the fact that
an ultra-high-energy photon would interact with a low-energy (``infrared'') photon of the CMN background (with energies in the eV range) to produce electron prositron pairs, according to the reaction:
\begin{equation}
  \gamma_{\rm UHE} \quad + \quad \gamma_{\rm CMB} \quad \Rightarrow \quad e^+~e^- ~.
\label{heirreact}
\end{equation}
The basic assumption in the analysis is the strict energy and momentum conservation in the
above reaction, despite the modified dispersion relations for the photons.
Such an assumption stems from the validity of the \emph{local-effective-lagrangian formalism}
to the case of an effective description of the QG foam effects on particles with energies much lower than the QG energy scale (assumed close to Planck scale $M_{\rm Pl} = 10^{19}$ GeV).
In this formalism, one can represent effectively the foam \emph{dispersive effects} as corresponding to higher-derivative \emph{local operators} in a \emph{flat-space-time} Lagrangian. The upshot of this is the modification of the pertinent equations of motion for the photon field (which in a Lorentz-invariant theory would be the ordinary Maxwell equations) by higher-derivative terms, suppressed by some power of the QG mass scale.

In the notation of \cite{sigl} one can consider the following modified dispersion relations:
\begin{eqnarray}\label{siglmdr}
  && \omega_{\pm} = k^2 + \xi_n^{\pm} k^2 \left(\frac{k}{M_{\rm Pl}}\right)^n~, \qquad \omega_b^2 = k_b^2~, \nonumber \\
 &&  E_{e,\pm}^2 = p_e^2 + m_e^2 + \eta^{e,\pm}_n p_e^2 \left(\frac{p_e}{M_{\rm Pl}}\right)^n~
\end{eqnarray}
where $(\omega, \vec k)$ indicate four-momenta for photons, $(E, \vec{p}_e $ are the corresponding four-momentum vectors for electrons, and the suffix $b$ indicates a low-energy CMB photon, whose dispersion relations are assumed approximately the normal ones, as any QG correction is negligible due to the low values of energy and momenta. The +(-) signs indicate left(right) polarizations (photons) or helicities (electrons). Positive (negative) $\xi$ indicate subluminal (superluminal) refractive indices.
Upon the assumption of energy-momentum conservation in the process,  one arrives at kinematic equations for the threshold of the reaction (\ref{heirreact}), that is the minimum energy of the high-energy photon required
to produce the electron-positron pairs.

For the linear- or quadratic- suppression case (Cases I and II in the MAGIC analysis above, for which $n=1,2$ respectively (\ref{siglmdr})), for instance, one finds that for the relevant subluminal photon refractive indices corresponding to the saturation of the $M_{QG1}$ lower bound in (\ref{mq12bound}), the threshold for pair production disappears for ultra-high-energy photons, and hence such photons should have been observed.
The non-observation of such photons implies constraints for the relevant parameters $\xi, \eta$
which are stronger by \emph{several orders of magnitude} than the bounds (\ref{mq12bound}) inferred from the MAGIC observations.

From the analysis of \cite{sigl} for the $n=1$ case, one concludes that in the case of linear Planck-mass suppression of the sub-luminal QG-indunced modified dispersion relations for photons, of interest for   the QG-foam interpretation of the MAGIC results~\cite{MAGIC2}, parameters with size $\xi_1 > 10^{14}$ are ruled out. This exceeds the sensitivity of the MAGIC experiment (c.f. (\ref{mq12bound})) to such Lorentz-symmetry violating effects by fifteen orders of magnitude !
Similar strong constraints are also obtained from the non observations of \emph{photon decay} ($\gamma \to e^+ e^-$), a process which, if there modified dispersion relations, is generally allowed~\cite{sigl}.

However, the reader should recall that this type of analysis is based on exact energy momentum conservation in the process (\ref{heirreact}), stemming from the assumption of the local-effective lagrangian formalism for QG-foam. As we discussed in \cite{emngzk}, and shall review briefly below, however, such a formalism need not be applicable in the case of quantum gravity, where the fluctuations of space-time or other defects of gravitational nature
paly the r\^ole of an external environment, resulting in \emph{energy fluctuations} in the reaction (\ref{heirreact}). The presence of such fluctuations does affect the relevant energy-threshold equations, for the reaction to occur, which stem from kinematics, in such a way that the above stringent limits are no longer valid.

\end{itemize}

 After this necessary digression, we now come back to discuss
 a string theory model (or better a class of such models), which can explain the MAGIC
 observations in terms of vacuum-induced non-trivial refractive indices, in agreement with all the
 above-mentioned complementary constraints on Lorentz Violations.

\section{A stringy model for space time medium with non-trivial ``optical'' properties: D--Particle ``foam''\label{sec:string}}

\begin{figure}[ht]\begin{center}
  \includegraphics[width=5cm]{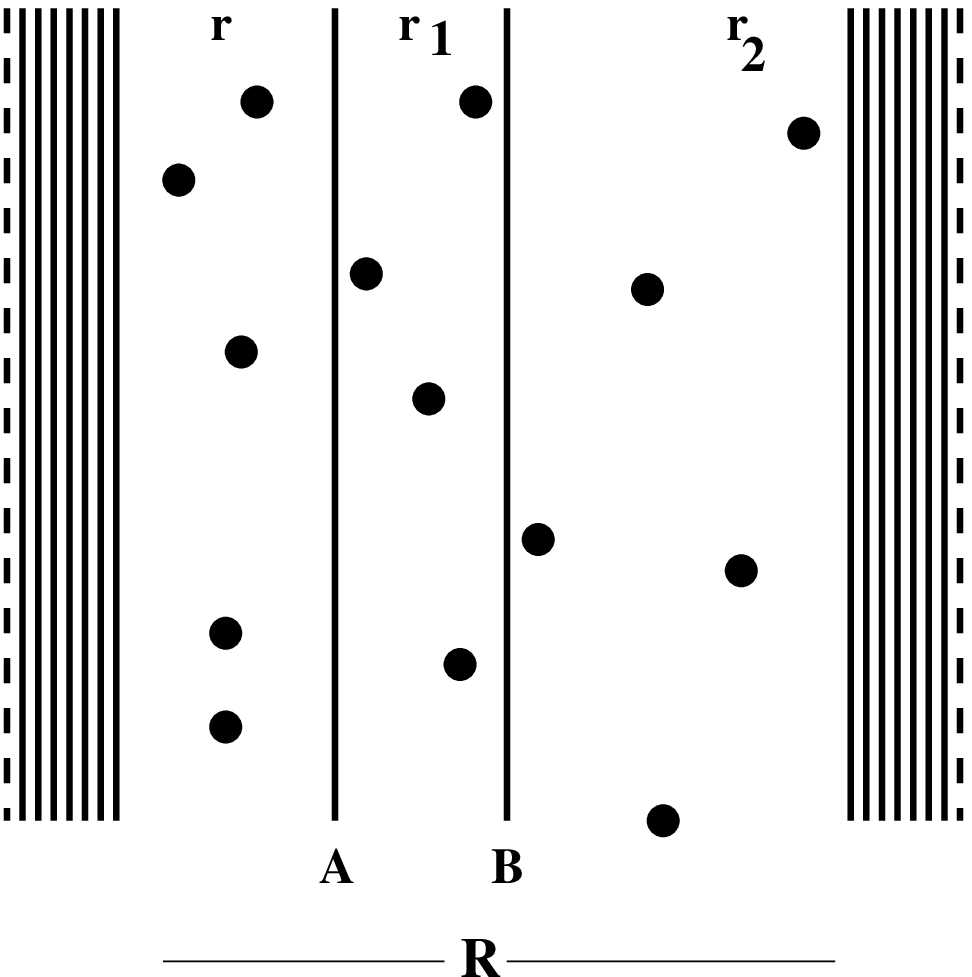} \hfill \includegraphics[width=5cm]{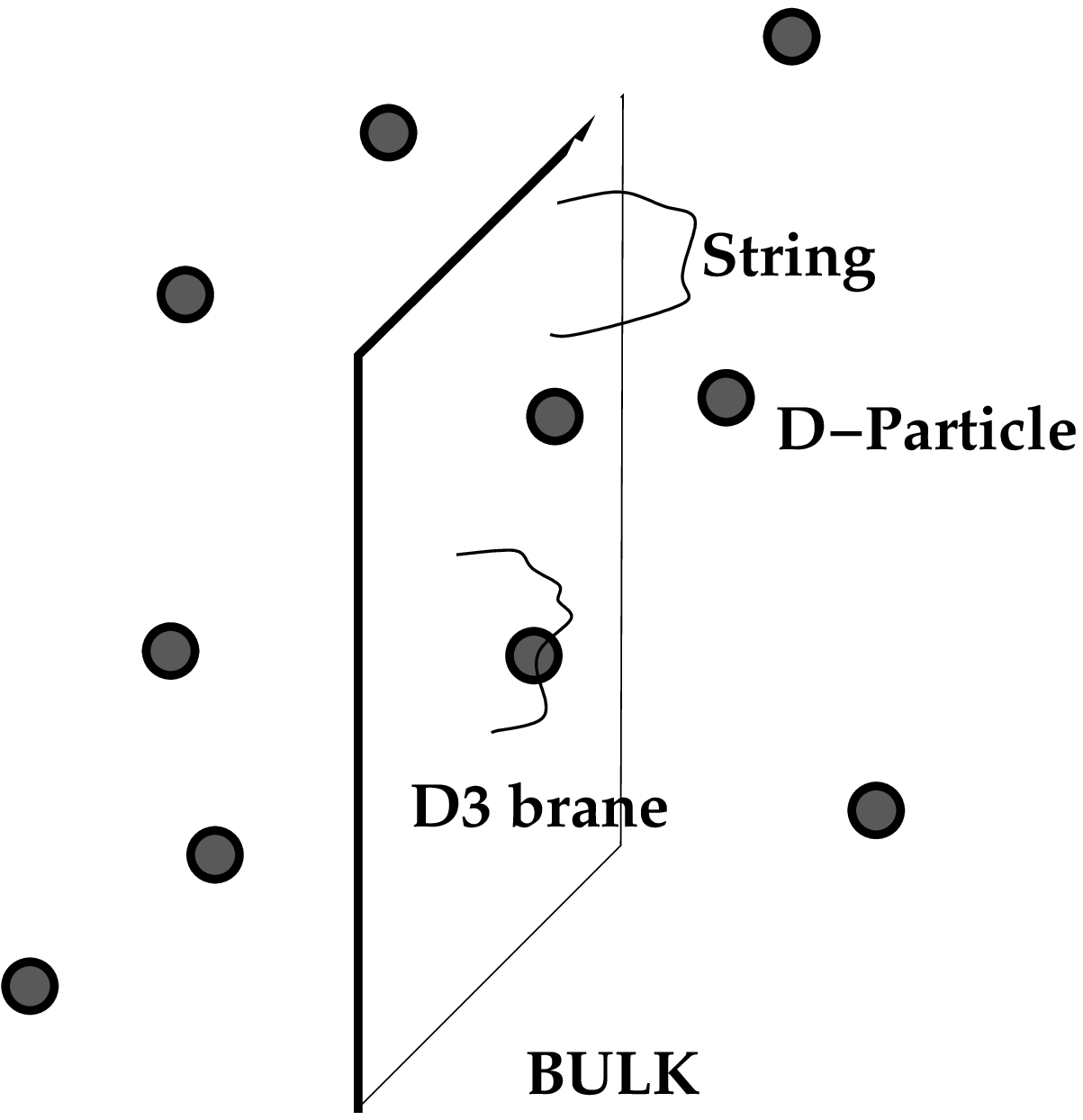}
\end{center}\caption{A type IA string theory model of D-particle ``foam''. The model consists of appropriate stucks (left panel) of D(irichlet)branes, some of which are moving in a higher-dimensional Bulk space time, punctured by point-like D-brane defects (D-particles). As the brane world moves (right panel), D-particles from the bulk cross the brane world, and  thus they appear to an observer on the brane as ``flashing on and off'' space-time foam defects (``D-particle foam''). Photons, represented as open strings on the D3 brane, interact with these defects via capture/recoil, and this leads to non-trivial refractive indices. The effect is therefore ``classical'' from the bulk space time point of view, but appears as an effective ``quantum foam'' from the D3-brane observer effective viewpoint.}
\label{dfoam:fig}
\end{figure}

From the above discussion it becomes clear that any model
of refraction in space-time foam that exhibits effects at the level of the MAGIC sensitivity~\cite{MAGIC2}  (\ref{mq12bound}) should be characterised by the following specific properties:
\begin{itemize}

\item{(i)} photons are \emph{stable} (i.e. do \emph{not} decay) but should exhibit a modified \emph{subluminal} dispersion relation with Lorentz-violating corrections that grow linearly with $E/(M_{\rm QG\gamma}c^2)$, where $M_{\rm QG\gamma}$ is close to the Planck scale,

\item{(ii)} the medium should not refract electrons, so as to avoid the synchrotron-radiation
constraints~\cite{crab,ems}, and

\item{(iii)} the coupling of the photons to the medium must be independent of photon polarization, so as not to have birefringence, thus avoiding the pertinent stringent constraints~\cite{uv,grb,crab2}.

\item{(iv)} The formalism of local effective lagrangians breaks down, in the sense that
there are quantum fluctuations in the total energy in particle interactions,
due to the presence of a quantum gravitational `environment', such that stringent constraints, which otherwise would be imposed from the non-observation of ultra-high energy photons ($\hbar \omega > 10^{19}$ eV), are evaded.
\end{itemize}

    A model with all these properties has been suggested by us some years ago~\cite{emnw,ems,emnnewuncert}, and is based on a stringy model for space-time foam, which we now come to discuss.

\subsection{Brief Review of the Model and its Cosmology}

In \cite{emnw}
we have attempted to construct a brane/string-inspired
model of space time foam which could have realistic cosmological properties.
For this purpose we exploited the modern approach to string theory~\cite{polch}, involving
membrane hypersurfaces (D(irichlet)-branes). Such structures are responsible for reconciliating (often via duality symmetries) certain string theories (like type I), which before were discarded as physically uninteresting, with Standard-Model phenomenology in the low-energy limit.

In particular, we
considered (c.f. figure \ref{dfoam:fig}) a ten-dimensional bulk bounded
by two eight-dimensional orientifold planes, which contains two stacks of
eight-dimensional branes, compactified to three spatial
dimensions. Owing to special reflective properties, the orbifolds act as
boundaries of the ninth-dimension. The bulk space is punctured by point-like D0-branes
(D-particles), which are allowed in type IA string theory (a T-dual of type I strings~\cite{schwarz}) we consider in \cite{emnw} and here~\footnote{One can extend the construction to phenomenologically realistic models~\cite{linano} of type IIB strings, in which the ``D-particles'' are constructed out of D3-branes wrapped around appropriate three cycles. The observable universe is represented by appropriate stacks of higher dimensional Dp-branes (with some of their extra dimensions compactified). Standard model excitations are represented by strings stretched among such branes. The foamy backgrounds of brane worlds punctured by D-particles consist of intersecting Dp-
D3(compactified)-brane configurations, with open strings stretching between them, which describe the capture process.
The calculation of time delays from such processes is technically more involved in such cases, but the basic conclusions remain the same as in our type-IA string model. Photons and not electrons are susceptible to significant time delays, leading to observed effects.}. These are massive objects in string theory~\cite{polch}, with masses
$M_s/g_s$, where $M_s$ is the string mass scale (playing the r\^ole of the quantum gravity scale in string theory), and $g_s < 1$ is the string coupling, assumed weak for our purposes. These
objects are viewed as space-time \emph{defects}, analogous, \emph{e.g.} to cosmic strings, but these are point-like and electrically neutral. I have to stress at this point that, according to modern ideas in string theory~\cite{polch}, the scale $M_s$ is in general different from the four-dimensional Planck-mass scale $M_P = 1.2 \times 10^{19}$ GeV/$c^2$, and in fact it is a free parameter in string theory to be constrained by experiment.
The energy scale $M_s c^2$ can be as low as a few TeV; it cannot be lower than this, though, since in such a case we should have already seen fundamental string structures experimentally, which is not the case.

Our model assumes a collision between two branes from the original stack of branes (c.f. figure \ref{dfoam:fig}) at an early
epoch of the Universe, resulting in an initial cosmically catastrophic
Big-Bang type event in such non-equilibrium cosmologies~\cite{brany}. After
the collision, the branes bounce back.

It is assumed that currently the
branes are moving slowly towards the stack of branes from which they
emanated. Hence populations of bulk D-particles cross the brane worlds and
interact with the stringy matter on them. To an observer on the brane the
space-time defects will appear to be {\em \textquotedblleft
flashing\textquotedblright } on and off. The model we are using involves
eight-dimensional branes and so requires an appropriate \emph{compactification}
scheme to three spatial dimensions e.g. by using manifolds with non-trivial
\emph{fluxes} (unrelated to real magnetic fields). Different coupling of fermions
and bosons to such external fields breaks target space supersymmetry. The
consequent induced mass splitting~\cite{bachas,gravanis} between fermionic
and bosonic excitations on the brane world is proportional to the intensity
of the flux field (a string generalization of the well-known Zeeman effect of ordinary quantum mechanics, whereby the presence of an external field leads to energy splittings, which are however different  between (charged) fermions and bosons.
In this way one may obtain phenomenologically realistic mass
splittings in the excitation spectrum (at TeV or higher energy scales) owing
to \emph{supersymmetry obstruction} rather than spontaneous breaking (this terminology, which is due to E.~Witten~\cite{obstr}, means that,
although the ground state could still be characterised by zero vacuum energy, the masses of fermion and boson excitations differ and thus supersymmetry is broken at the level of the excitation-spectrum).
The assumption
of a population concentration of
massive D-particle defects in the haloes of galaxies can lead to modified galactic dynamics~%
\cite{recoil2}. Hence, we have an alternative scenario to standard cold dark
matter, using vector instabilities arising from the splitting of strings and
attachment of their free ends to a D-particle defect (c.f. fig.~\ref{dfoam:fig}).

\begin{figure}[ht]
 \begin{center} \includegraphics[width=7cm]{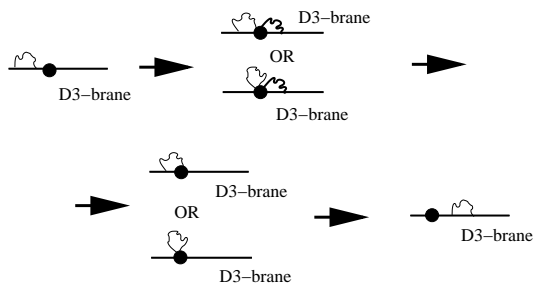}
\end{center}\caption{Schematic view of the capture process of an open string state, representing a photon, propagating on a D3-brane world by a D-particle on this world. The intermediate string state, indicated by thick wave lines, which is creating on capture of the end(s) of the photon by the D-particle, stretches between the D-particle and the brane world, oscillates in size between $0$ and $\alpha ' p^0$, where $p^0$ is the energy of the incident photon, and thus produces a series of outgoing photon waves, with attenuating amplitudes, constituting the re-emission process. The intermediate string state provides also the restoring force, necessary for keeping the D-particle roughly in its position after scattering with the photon string state.}
\label{fig:restoring}
\end{figure}

\subsection{Time Delays in D-particle foam: a matter of Uncertainty with...strings attached \label{sec:uncert}}

We can now proceed to discuss a possible origin of time delays induced in the arrival time of photons, emitted simultaneously from an astrophysical object, as a result of their propagation
in the above-described D-particle space-time foam model.
As we shall argue, the above model is in principle capable of reproducing photon arrival time delays proportional to the photon energies, of the kind observed in the MAGIC experiment~\cite{MAGIC,MAGIC2}. Such delays occur during the encounter of photons with the D-particles defects in the foam. This microscopic phenomenon, which -as we shall see below-is essentially stringy and does not \emph{characterise} local field theories, contributes to a sub-luminal non-trivial
refractive index \emph{in vacuo}, induced by the capture of photons or electrically neutral probes by the D-particle
foam~\cite{emnnewuncert}.
The capture process is described schematically in figure \ref{fig:restoring} and we next proceed to describe the underlying physics, which is essentially \emph{stringy}.

An important feature of the model is that, on account of electric charge conservation, only electrically neutral excitations are subjected to capture by the D-particles. To the charged matter the D-particle foam looks \emph{transparent}. This is because the capture process (\ref{fig:restoring}) entails a splitting of the open string state, representing matter excitations. Charged excitations are characterized by an electric flux flowing across the string, and when the latter is cut in two pieces as a result of its
capture by the D-particle defect, the flux should go somewhere because charge is conserved.
The D-particle, being neutral, cannot support this conservation, and hence only \emph{electrically neutral} excitations, such as photons, are subjected to this splitting and the associated delays, as we shall discuss below. The reader should bear in mind of course that the D-particles carry other kinds of fluxes, unrelated to electromagnetism~\cite{polch}. These are conserved separately, and it is for this reason that isolated D-particles cannot exist, but there must always be in the company of other D-branes, as in our model above, so that the relevant fluxes are carried by the stretched strings between the latter and the D-particles.

For our purposes in this work it is also important to remark that the D-particles are treated as static when compared to photons. This is because the ends of the open string representing the photon move on the D3 brane world with the speed of light in (normal) vacuo, $c=1$ in our units, while the relative velocities of the D-particles with respect to the brane world (which propagates in the bulk space) are much lower than this, in the model of brane cosmology of \cite{emnw}, which we use here as our concrete example of D-particle foam. For instance, as discussed in \cite{brany}, to reproduce cosmological observations in this model, in particular the spectrum of primordial density fluctuations, which are affected by the relative motion of the brane world, the speed of propagation of the D3-brane Universe should be smaller than $10^{-4}c$.

When the end(s) of the open-string photon state are attached to the D-particle, there is an intermediate string state formed, \emph{stretched} between the D-particle and the D3-brane, which absorbs the incident energy $p^0 c$ of the photon state (where $p^0$ denotes the temporal component of the four-momentum), to grow in size from zero size to a length $L$ that is determined by the requirement of energy minimization as follows. One assumes~\cite{sussk2} that the intermediate string state needs $N$ oscillations to achieve its maximal length $L$. This implies:
\begin{equation}
   p^0 = \frac{L}{\alpha '} + \frac{N}{L}
\label{noscill}
\end{equation}
where $\alpha ' = \ell_s^2$ is the Regge slope, which is equal to the square of the string length $\ell_s$,
and hence it has dimensions of [length]$^2$. In (\ref{noscill}) and throughout the article we are working in relativistic units of $\hbar = c=1$, and in this sense $p^0$ is identified with the energy of the photon. In such units we also have that the string length $\ell_s = 1/M_s$, where $M_s$ is the string mass scale.

The first term on the right-hand side of (\ref{noscill}), proportional to the
 length $L$, is due to the tension $1/\alpha '$ of the stretched string, while the second term represents the energy storage due to oscillations. The relation (\ref{noscill}) then expresses energy conservation
 in the capture process of fig.~\ref{fig:restoring}.

We can minimize the right hand side by varying with respect to $L$ and demanding the derivative to vanish, which guarantees energy minimization in the process. This procedure determines $N$, which then is substituted back to the equation to yield the required maximal $L$: $L_{\rm max} = \frac{1}{2}\alpha ' p^0$. Since the end of the stretched string state attached on the D3-brane moves  with the sped of light in (normal) vacuo $c=1$, this implies that the time taken (delay) for the intermediate string state to first grow to this maximal length and then shrink again to zero size is:
\begin{equation}
  \Delta t \sim  \alpha ' p^0
\label{delayonecapture}
\end{equation}
This describes the time delay encountered in photon propagation in D-particle foam, due to the formation of the intermediate composite string state between D-particles and the photon (fig. \ref{fig:restoring}).

Above we have discussed the situation in the Dirichlet picture, describing the attachment of the end of the strings on D-branes.  The (quantum) oscillations of the intermediate string state will produce a series of outgoing wave-packets, with attenuating amplitudes, which will correspond to the re-emission process of the photon after capture. The presence of the stretched string state, which carries the characteristic flux of the D-brane interactions, provides the \emph{restoring force}, necessary to keep the D-particle in its position after scattering with the photon. The situation may be thought of as the stringy/brany analogue of the restoring force in situations in local field theories of photons propagating in media with non trivial refractive indices, as discussed by Feynman~\cite{feynman} and reviewed above in section (\ref{sec:ntv}). The r\^ole of the electrons in that case (represented as harmonic oscillators) is played here by the D-particle defects of the space-time. The stringy situation, however, is more complicated, since the D-particles have an infinite number of oscillatory excitations, represented by the various modes of open strings with their ends attached to them. Moreover, contrary to the conventional medium case, in the string model the refractive index is found proportional to the photon frequency, while the effective mass scale that suppresses the effect (\ref{delayonecapture}) is the quantum gravity (string) scale $M_s$ and not the electron mass, as in (\ref{refrordinary}). The latter property can be understood qualitatively by the fact that the mass of these D-particle defects is of order~\cite{polch} $1/(g_s~\sqrt{\alpha '}) = M_s/g_s$, where $g_s$ is the string coupling (in units of $\hbar = c = 1$).

The time delays (\ref{delayonecapture}) pertain to a single encounter of a photon with a D-particle. In case of a foam, with a linear density of defects $n^*/\sqrt{\alpha '}$, \emph{i.e.} $n^*$ defects per string length, the overall delay encountered in the propagation of the photon from the source to observation, corresponding to a traversed distance $D$, is:
\begin{equation}
\Delta t_{\rm total} = \alpha ' p^0 n^* \frac{D}{\sqrt{\alpha '}} = \frac{p^0}{M_s} n^* D
\label{totaldelay}
\end{equation}
When the Universe's expansion is taken into account, one has to consider the appropriate red-shift-$z$ dependent stretching factors which affect the measured delay in the propagation of two photons with different energies.
Thus, from (\ref{totaldelay}) we obtain in such a case a total delay in the arrival times of photons with energy difference $\Delta E$,  which has the form considered in \cite{mitsou}, namely it is proportional to $\Delta E$ and is suppressed linearly by the quantum gravity (string) scale, $M_s$:
\begin{equation}
(\Delta t)_{\rm obs} =  \frac{ \Delta E}{M_s} {\rm H}_0^{-1}\int_0^z n^*(z) \frac{(1 + z)dz}{\sqrt{\Omega_\Lambda + \Omega_m (1 + z)^3}}
\label{redshift}
\end{equation}
where $z$ is the red-shift, ${\rm H}_0$ is the (current-era) Hubble expansion rate, and we have
assumed for concreteness the $\Lambda$CDM standard model of cosmology, with
$\Omega_i \equiv \frac{\rho_{i(0)}}{\rho_c}$ representing the present-epoch energy densities ($\rho_i$) of matter (including dark matter), $\Omega_m$, and dark vacuum energy, $\Omega_\Lambda$, in units of the critical density $\rho_c \equiv \frac{3H_0^2}{8\pi G_N}$ of the Universe ($G_N$ is the Newton's gravitational constant), that is the density required so that the Universe is spatially flat. The current astrophysical measurements of the
acceleration of the Universe are all consistent with a non zero Cosmological-Constant Universe with Cold-Dark-Matter ($\Lambda$CDM Model), with $\Omega_\Lambda \sim 73 \%$ and $\Omega_m \sim 27\% $.

Notice in (\ref{redshift}) that the essentially stringy nature of the delay implies that the characteristic suppression scale is the string scale $M_s$, which plays the r\^ole of the quantum gravity scale in this case. The scale $M_s$ is a free parameter in the modern version of string theory, and thus it can be constrained by experiment. As we have discussed in this article, the observations of delays of energetic (TeV) photons from AGN by the MAGIC telescope~\cite{MAGIC} can provide such an experimental way of constraining $n^*/M_s$ in (\ref{redshift}). For $\Delta E \sim 10$ TeV, for instance, the delay (\ref{redshift}) can lead to the observed one of order of minutes, provided $M_s/n^* \sim 10^{18}$ GeV (in natural units with $c=1$)~\cite{MAGIC2}. This implies natural values for both $n^*$ and $M_s$, although it must be noted that $n^*$ is another free parameter of the bulk string cosmology model of \cite{emnw}, considered here. In general, $n^*(z)$ is affected by the expansion of the Universe, as it is diluted by it. This depends on the bulk model and the interactions among the D-particles themselves. For redshifts of relevance to the MAGIC experiments, $z=0.034 \ll 1$, one may ignore the $z$-dependence of $n^*$ to a good approximation.

The total delay (\ref{redshift}) may be thought of as implying~\cite{feynman} an effective \emph{subluminal} refractive index $n(E)$ of light propagating in this space time, since one may assume that the delay is equivalent to light being slowed down due to the medium effects.
 On account of the theoretical uncertainties in the source mechanism, however, the result of the
 AGN Mkn 501 observations of the MAGIC Telescope translate to \emph{upper} bounds for the quantity $n^*/M_s$ in (\ref{redshift}), which determines the strength of the anomalous photon dispersion in the string/D-particle foam model.

We remark at this stage that the above time delays are a direct consequence of the \emph{stringy uncertainty principles}, and as we have discussed they are essentially stringy effects, associated with the capture process of fig. \ref{fig:restoring}. Indeed, strings are characterised -- apart from the phase-space Heisenberg uncertainty relation, modified by higher order terms in $\alpha '$~\cite{venheisenberg} as a result of the existence of the minimal string length $\ell_s \equiv \sqrt{\alpha '}$ in target space-time,
\begin{equation}
  \Delta X \Delta P \ge \hbar  + \alpha ' (\Delta P)^2 + \dots
\label{heisenberg}
\end{equation}
also by the space-time uncertainty relation~\cite{yoneya}
Such space-time uncertainty relations are consistent with the corresponding space-time string uncertainty principle~\cite{yoneya}
\begin{equation}
   \Delta X \Delta t \ge \alpha '
\label{stringyunc}
\end{equation}
Since the momentum uncertainty $\Delta P < p^0$, we have from (\ref{heisenberg}), to leading order in $\alpha '$ (in units $\hbar = 1$):
\begin{equation}
\Delta X \ge \frac{1}{\Delta P} > \frac{1}{p^0}
\end{equation}
In view of (\ref{delayonecapture}), we then arrive at consistency with the space-time un certainty (\ref{stringyunc}),
\begin{equation}
\Delta X \ge \frac{\alpha'}{\alpha ' p^0} \sim \frac{\alpha '}{\Delta t}
\label{timespacenew}\end{equation}

These delays are \emph{causal}, i.e. consistent with the fact that signals never arrive before they occur.
They do not characterize local field theories in non commutative space times (the analogue of having strings in constant electric fields), which suffer from \emph{non causal effects}, due to the existence of advanced outgoing waves after the scattering. For the expert reader we note that
the fact that string theory produces only retarded waves after the scattering, and not advanced ones,
has been explained by a detailed calculation in \cite{sussk2} in the case of the scattering of two open-string modes (scalar tachyon modes, for simplicity). There are scattering phases which are such that in string theory only delays occur. This issue is related with the \emph{maintenance of causality} by strings, which notably is violated in the case of field theories in non-commutative space times, where a similar scattering of wave packets would result to advanced outgoing wave-packets, violating causality.

In view of the above discussion, if the time delays observed by MAGIC can finally be attributed partly or wholly to this type of stringy space-time foam, then the AGN Mkn 501, and other such celestial sources of very high energy photons, may be viewed as playing the r\^ole of the ``Heisenberg microscopes'' for these string/brane  uncertainties.

\section{Instead of Conclusions \label{sec:5}}

From the above discussion it becomes clear that the properties/requirements (i)-(iv)
at the beginning of the last section (\ref{sec:string}),
 which the photon must satisfy, so that the MAGIC results on the time four-minute delay~\cite{MAGIC2} are attributed to propagation in a stochastic quantum-gravity medium, are indeed respected by the
 D-particle string foam model just presented.
 This of course does not mean that there are no conventional astrophysics explanations for the MAGIC results, but it demonstrates clearly that  string theory (or better its modern version involving D-brane defects) is capable of explaining the observed photon delays, in agreement with all the other astrophysical data currently available.
 This is at least amusing, since it provides a framework for experimentally testing some models of string theory at present or in the foreseeable future.

 The key point of course in the approach is the existence of space-time defects in the ground state of the model, whose topologically non-trivial interactions with the
 string states, via string-stretching during the capture process (c.f. fig.~\ref{fig:restoring}),
 are mainly responsible for the observed delays, proportional to the incident photon energy.
The peculiarity of the D-particle foam in being \emph{transparent} to charged particles (as a result of electric charge conservation requirements), evades the stringent constraints on linear Planck scale suppression refractive indices that would be induced by electron synchrotron radiation studies from Crab Nebula~\cite{crab}. Moreover, the absence of birefringence avoids the similarly stringent constraints on such models that otherwise would be imposed by galactic measurements~\cite{crab2}.
Finally, it worths mentioning that the string-stretched linear in energy time delays (\ref{delayonecapture}),(\ref{redshift}), when applied to neutrinos, can be flavour (i.e. neutrino-species) independent, thus avoiding~\cite{ellis08} the stringent constraints that would be obtained from models of quantum gravity with flavour-dependent modifications of neutrino propagation and thus modifications in their oscillations~\cite{brustein}.

As already mentioned, the MAGIC result needs confirmation by other experiments like H.E.S.S.~\cite{hess2155,hessnew}, or other photon dedicated experiments, like FERMI (formerly GLAST)~\cite{glast}, where photons from Gamma Ray bursts will be observed.
One needs many more high-energy astrophysical photon measurements to be able to disentangle source from possible propagation effects due to fundamental physics.
If a statistically significant population of data on photons from cosmic sources is collected, exhibiting refractive indices varying linearly with the distance of the source~\cite{mitsou}, as well as the photon energy, then this would be a very strong confirmation of the D-particle foam model, for reasons explained above. However, it must be noted that GRB's, which are expected to lead to statistically significant data in the next few years, will produce photons much lower in energies than the flares observed in AGN, and this could be a drawback. At any rate there are attempts to claim that observations from FERMI will have sensitivity close to the Planck scale~\cite{lamon} for such linear-suppression models. And, indeed, as we mentioned above, in September 2008 the FERMI collaboration claims~\cite{grbglast} observations of  a delayed arrival of high energy ($\sim$ 13 GeV) photons, as compared to lower-energy ones, from the distant GRB 080916c, which can be fitted excellently by the above-described string theory model of space time foam, with a string scale of order $10^{18}$ GeV, in agreement with the MAGIC results. The situation starts to becoming exciting...

From the above discussion it becomes, hopefully, clear to the reader that experimental searches for quantum gravity, if the latter is viewed as a medium, are highly model dependent as far as the sensitivity
to experimental falsification of the predictions of the underlying theoretical models is concerned. However, we are entering an era where low-energy (compared to Planck scale) physics experiments can already provide valuable information on the structure of space-time at the scales where Quantum Gravity is expected to set in. Very- and Ultra- high energy Astrophysics is at the forefront of such fundamental research. We therefore expect that, for the years to come, this branch of physics will proceed in parallel with terrestrial high energy experiments, such as the Large Hadron Collider launched at CERN recently, and be able to provide us soon with complementary
important information on the underlying fundamental structure of our Cosmos.
Time will then show whether quantum gravity can be finally put to experimental confirmation. ~\emph{Affaire \`a suivre...}

\section*{Acknowledgements}

It is a real pleasure to thank V.A. Mitsou for discussions, especially on astro-particle physics issues.
I also wish to thank the organisers of the DICE 2008 Conference (Castiglioncello, Italy) for the invitation to speak, and for creating a
stimulating atmosphere for discussions during the meeting.
This work  is partially supported by the European Union
through the Marie Curie Research and Training Network \emph{UniverseNet}
(MRTN-2006-035863).

\end{document}